 \documentclass[11pt,reqno]{amsart}
 \usepackage{amsxtra}
 \usepackage[innercaption]{sidecap}
 \usepackage{graphicx}
 \graphicspath{{./images/}}
 \usepackage{amssymb}
 \usepackage{enumitem}

\theoremstyle{plain}

\def\CC {{\mathbb C}}
\def\RR {{\mathbb R}}
\def\NN {{\mathbb N}}
\def\ZZ {{\mathbb Z}}
\def\PP {{\mathbb P}}

\def\be {\begin{eqnarray}}
\def\ben {\begin{eqnarray*}}
\def\ee {\end{eqnarray}}
\def\een {\end{eqnarray*}}

\def\ds  {\displaystyle}

\def\AAA{\kern-0.3em}
\def\AA{\kern-0.18em}
\def\AC{\kern-0.14em}
\def\AB{\kern-0.22em}

\newcommand \nc {\newcommand}

\newtheorem{theorem}{Theorem}[section]
\newtheorem{lemma}[theorem]{Lemma}
\newtheorem{proposition}[theorem]{Proposition}
\newtheorem{corollary}[theorem]{Corollary}
\newtheorem{definition}[theorem]{Definition}
\newtheorem{example}[theorem]{Example}
\newtheorem{remark}[theorem]{Remark}
\nc \bth[1] { \begin{theorem}\label{t#1} } \nc \ble[1] {
\begin{lemma}\label{l#1} } \nc \bpr[1] {
\begin{proposition}\label{p#1} } \nc \bco[1] {
\begin{corollary}\label{c#1} } \nc \bde[1] {
\begin{definition}\label{d#1}\rm } \nc \bex[1] {
\begin{example}\label{e#1}\rm } \nc \bre[1] {
\begin{remark}\label{r#1}\rm } \nc \bcon[1] {
\medskip\noindent{\it{Conjecture #1}} } \nc \bqu[1]  {
\medskip\noindent{\it{Question #1}} }
\nc {\ethe} { \end{theorem} }
 \nc {\ele} { \end{lemma} } \nc {\epr}
{ \end{proposition} } \nc {\eco} { \end{corollary} } \nc {\ede} {
\end{definition} } \nc {\eex} { \end{example} } \nc {\ere} {
\end{remark} } \nc {\econ} {\smallskip} \nc {\equ} {\smallskip}
 \nc \thref[1]{Theorem \ref{t#1}}
\nc \leref[1]{Lemma \ref{l#1}} \nc \prref[1]{Proposition
\ref{p#1}} \nc \coref[1]{Corollary \ref{c#1}} \nc
\deref[1]{Definition \ref{d#1}} \nc \exref[1]{Example \ref{e#1}}
\nc \reref[1]{Remark \ref{r#1}}
\def \B {{\mathcal B}}
\def \T {{\mathcal T}}
\def \L {{\mathcal L}}

\def \diag { {\mathrm{diag}} }

 \def\AA  {\kern-0.1em}
 \def\BB  {\kern+0.1em}
 \def\BBB {\kern+0.15em}
 \def\K   {\kern+0.05em}
 \def\MK  {\kern-0.07em}
 \def\MKK {\kern-0.04em}
 \def\ds  {\displaystyle}

\begin{document}

\vspace{0.5cm}

\title[ Non-integrability of the Sasano system ]
{ Non-integrability of the  Sasano system of type  $A_5^{(2)}$ }

\author[Tsvetana  Stoyanova]{Tsvetana  Stoyanova}

\date{01.12.2025}

 \maketitle

\begin{center}
{Faculty of Mathematics and Informatics,
Sofia University "St. Kliment Ohridski",\\ 5 J. Bourchier Blvd., Sofia 1164, Bulgaria,
cveti@fmi.uni-sofia.bg}
\end{center}

\vspace{0.5cm}

\begin{abstract}
    The Sasano system of type $A^{(2)}_5$ is a four-dimensional  non-linear system of ordinary differential equations,
which has an affine Weyl group of  symmetries of type $A^{(2)}_5$.
 It is also a time dependent Hamiltonian system, which can be considered as coupled Painlev\'e III systems. In this paper,
utilizing the Morales-Ramis-Sim\'o theory for integrability of Hamiltonian systems, we prove rigorously that for all values of the parameters,
 for which the Sasano system of type $A^{(2)}_5$ admits a particular rational solution, it is non-integrable by rational first integrals.
	 \end{abstract}

   \maketitle



\headsep 10mm \oddsidemargin 0in \evensidemargin 0in

\section{Introduction}

In a series of papers \cite{Sa, Sa1, Sa2, Sa3, Sa4, Sa5, Sa6} Yusuke Sasano has introduced higher order nonlinear systems of ordinary differential equations, which have certain affine Weyl groups of symmetries
 as groups of B\"acklund transformations. All of these systems, which are called the Sasano systems, are time dependent Hamiltonian systems. The four-dimensional
Sasano systems can be considered as coupled Painlev\'e systems.
In this paper we study the integrability of the Sasano system of type $A^{(2)}_5$ given by
  \be\label{s}
   \dot{x} &=&
   \frac{2\,x^2\,y - x^2  +( \alpha_0+\alpha_1+\alpha_3)\,x}{t} - 1 + 4\,w +  \frac{2\,x\,z\,w}{t},\nonumber\\[0.15ex]
    \dot{y} &=&
    \frac{-2\,x\,y^2 + 2\,x\,y - (\alpha_0+\alpha_1+\alpha_3)\,y +\alpha_0}{t} - \frac{2\,y\,z\,w}{t},\\[0.15ex]
   \dot{z} &=&
   \frac{2\,z^2\,w - z^2 +(\alpha_0+\alpha_1+\alpha_3)\,z}{t} -1 + 4\,y  + \frac{2\,x\,y\,z}{t},\nonumber\\[0.15ex]
    \dot{w} &=&
    \frac{-2\, z\, w ^2 + 2 \,z\,w - (\alpha_0+\alpha_1+\alpha_3)\,w + \alpha_1}{t} - \frac{2\,x \,y\,w}{t}\,,\nonumber
  \ee
 where  $\alpha_j, 0 \leq j \leq 3$
 are complex parameters which satisfy the  relation
   \be\label{r}
    \alpha_0+ \alpha_1 + 2 \alpha_2 + \alpha_3=\frac{1}{2}\,.
    \ee
  The Sasano system \eqref{s} admits  a group of B\"acklund transformations \cite{Sa1}, which are defined by
        \be\label{sym}
                  s_0 : & &
           (*) \rightarrow \left(x + \frac{\alpha_0}{y}, y, z, w, t; -\alpha_0, \alpha_1, \alpha_2+\alpha_0, \alpha_3\right),\nonumber\\[0.3ex]
                  s_1 : & &
           (*) \rightarrow \left(x, y, z + \frac{\alpha_1}{w}, w, t; \alpha_0, -\alpha_1, \alpha_2+\alpha_1, \alpha_3\right),\\[0.3ex]
                  s_2 : & &
           (*) \rightarrow \left(x, y - \frac{\alpha_2\,z}{x\,z+t}, z, w + \frac{\alpha_2\,x}{x\,z +t}, t; \alpha_0+\alpha_2, \alpha_1+\alpha_2, -\alpha_2, \alpha_3 + 2 \alpha_2\right),\nonumber\\[0.3ex]
                  s_3 : & &
           (*) \rightarrow \left(x + \frac{\alpha_3}{y+w-1}, y, z + \frac{\alpha_3}{y+w-1}, w, t; \alpha_0, \alpha_1, \alpha_2+\alpha_3, -\alpha_3\right),\nonumber\\[0.3ex]
                   \pi : & &
           (*) \rightarrow \left(z, w, x, y, t; \alpha_1, \alpha_0, \alpha_2, \alpha_3\right)\,,\nonumber
                              \ee
              where we denote     $(*):=(x, y, z, w, t; \alpha_0, \alpha_1, \alpha_2, \alpha_3)$.
       The transformations \eqref{sym} satisfy the relations
          \ben
                      & &
           s_i^2=1, \quad i=0, 1, 2, 3, \qquad (s_0\,s_1)^2=(s_0\,s_3)^2=(s_1\,s_3)^2=1,\\
                     & &
                     (s_0\,s_2)^3=(s_1\,s_2)^3=1,\qquad (s_2\,s_3)^4=1, \qquad \pi^2=1\,.
             \een
      Thus $s_i, i=0, 1, 2, 3$ and $\pi_0$   generate the extended affine Weyl group $\widetilde{W}(A^{(2)}_5)$.

  The  system \eqref{s} is a time-dependent  Hamiltonian system of $2+1/2$ degrees of freedom
   \ben
    \frac{d x}{d t}=\frac{\partial H}{\partial y},\quad
   \frac{d y}{d t}=-\frac{\partial H}{\partial x},\quad
    \frac{d z}{d t}=\frac{\partial H}{\partial w},\quad
    \frac{d w}{d t}=-\frac{\partial H}{\partial z}\,,
   \een
   with the polynomial Hamiltonian
\begin{multline*}
                    H= \frac{1}{t} \Big[y\,(y-1)\,x^2 + (\alpha_0 + \alpha_1 + \alpha_3)\,x\,y - \alpha_0\,x - t\,y + \\[0.2ex]
                       +w\,(w-1)\,z^2 + (\alpha_0+\alpha_1 + \alpha_3)\,z\,w - \alpha_1\,z - t\,w + 4 t\,y\,w +2 x\, y\,z\,w \Big]=\\[0.2ex]
                    = H_{III}(x, y, t; \alpha_0 + \alpha_1+\alpha_3, \alpha_0) + H_{III}(z, w, t ; \alpha_0+\alpha_1+\alpha_3, \alpha_1) + \\[0.2ex]
                         + 4 y\,w + \frac{2 x\,y\,z\,w}{t},
                  \end{multline*}
                   where $H_{III}$ is the Hamiltonian associated with the third Painlev\'e equation given in the form
                  $$
                        H_{III}(q, p, t; a_1, a_2)=\frac{1}{t}\left[
                           p\,(p-1)\,q^2 + (a_1+a_2)\,q\,p - t\,p - a_2\,q\right].
                 $$

             In this way the Sasano system \eqref{s} can be considered as a coupled Painlev\'{e} III system.
      The system \eqref{s} can be turned into an autonomous Hamiltonian system with three degrees of freedom by introducing two new dynamical variables : $t$ and a conjugate
   to it $F$. The new Hamiltonian becomes
                    $$\,
                         \widehat{H}=H+F
                      \,$$
    and the associated Hamiltonian system is
        \be\label{H}
                               & &
    \frac{d x}{d s}=\frac{\partial \widehat{H}}{\partial y},\quad
    \frac{d y}{d s}=-\frac{\partial \widehat{H}}{\partial x},\quad
    \frac{d z}{d s}=\frac{\partial \widehat{H}}{\partial w},\quad
    \frac{d w}{d s}=-\frac{\partial \widehat{H}}{\partial z},\\[0,25ex]
                             & &
      \frac{d t}{d s}=\frac{\partial \widehat{H}}{\partial F},\quad
    \frac{d F}{d s}=-\frac{\partial \widehat{H}}{\partial t}\,.\nonumber
   \ee
              The symplectic structure $\Omega$ is canonical in the variables $(x, y, z, w, t, F)$, that is
         $\Omega=d y \wedge d x + d w \wedge d z + d F \wedge d t$.

       In this paper we will prove that for all values of the parameters, for which the Sasano system \eqref{s} admits a particular rational solution, it is non-integrable by rational first integrals.
            Recall that from the theorem of Liouville-Arnold \cite{Ar} this means the non-existence of
     three    functionally independent first integrals.
This article falls in the frame of the applications of the Morales-Ramis-Sim\'o theory to the integrability of the Painlev\'e systems and their higher-order analogues.
           Roughly speaking, depending on the approach, there are three basic school studying such kind of problems for the Painlev\'e systems or equations.
       The Japanese school, in the person of Umemura, Nishioka,  Noumi and Okamoto provides the first rigorous proofs of  the irreducibility of solutions of the Painlev\'e 
      equations utilizing nonlinear finite-dimensional  differential Galois theory \cite{N, NO, U}.
           The French school, in the person of Casale, Weil, Cantat and Loray  proves the  irreducibility of the Painlev\'e equations, not of their solutions, applying the theory on  Malgrange groupoid \cite{CL, C1, CW} .
         The third school, to which the author belongs, proves the non-integrability of the Painlev\'e systems and their higher order analogues utilizing the Morales-Ramis-Sim\'o theory for the integrability of
       Hamiltonian systems \cite{CG, F, HS, M, St1, St2, St3, St4, St5}. The Morales-Ramis-Sim\'o theory transforms the problem of integrability of a given
       analytic Hamiltonian system  to the problem of integrability of the variational equations along a  non-equilibrium particular solution.
   Since the variational equations are linear ordinary differential equations, their integrability is well defined in the context of the differential Galois theory.
  With the present paper  we continue our study on the integrability of the Sasano systems. 
         In \cite{St4} we have proved that for all values of the parameters, for which the Sasano system of type $A^{(2)}_4$ has a particular rational solution,
     it is non-integrable by rational first integrals. In \cite{St5} we have provided a non-integrable result for a one-parametric family of the Sasano system of type
       $D^{(2)}_5$.

     In \cite{Mat} Matsuda classify all the rational solutions of the Sasano system of type $A^{(2)}_5$. More precisely,
              \bth{M} {\bf (Matsuda)}
            For a rational solution of the Sasano system \eqref{s}, by some B\"acklund transformations, the solution and the parameters can be transformed so that
            \be\label{seed}
                         & &
             (x, y, z, w)=0, 1/4, 0, 1/4) \quad \textrm{and}\\
                          & &
            (\alpha_0, \alpha_1, \alpha_2, \alpha_3)=(\alpha_3/2, \alpha_3/2, \alpha_2, \alpha_3)=(\alpha_3/2, \alpha_3/2, 1/4-\alpha_3, \alpha_3), \nonumber
             \ee
                   respectively. Furthermore, for the Sasano system \eqref{s}, there exists a rational solution if and only if one of the following occurs:

                              \ben
                                -2 \alpha_0 + \alpha_3\in\ZZ,   & &  -2 \alpha_1+\alpha_3\in\ZZ,\\[0.1ex]
                                -2 \alpha_0 + \alpha_3\in\ZZ,     & & 2 \alpha_1+\alpha_3\in\ZZ,\\[0.1ex]
                                 2 \alpha_0 + \alpha_3\in\ZZ,   & &  -2 \alpha_1+\alpha_3\in\ZZ,\\[0.1ex]
                                 2 \alpha_0 + \alpha_3\in\ZZ,     & & 2 \alpha_1+\alpha_3\in\ZZ,\\[0.1ex]
                                -2 \alpha_0 + \alpha_3\in\ZZ,   & &  \alpha_3-1/2\in\ZZ,\\[0.1ex]
                                -2 \alpha_1 + \alpha_3\in\ZZ,     & & 2 \alpha_3-1/2\in\ZZ\,.
                            \een
                \ethe
            From \thref{M} it follows that all rational solutions of the Sasano system of type $A^{(2)}_5$ can be obtained from the seed solution \eqref{seed} by using the B\"acklund transformations \eqref{sym}.
              Studying the first normal variational equations along the seed solution \eqref{seed} we obtain the first key non-integrable result of this paper.

  \bth{key1}
    Assume that $\alpha_0=\alpha_1=\frac{\alpha_3}{2},\,\alpha_2=\frac{1}{4}-\alpha_3$, where $\alpha_3\in\CC$ is arbitrary, such that
   $4 \alpha_3-\frac{1}{2} \notin \ZZ$. Then the Sasano system \eqref{s} is not integrable by rational first integrals.
 \ethe

   In the case  when $4 \alpha_3-\frac{1}{2} \in \ZZ$, using only the Morales-Ramis theorem, we cannot say if the Sasano system \eqref{s} is integrable or non-integrable,
   since the connected component of the unit element of the differential Galois group of the first normal variational equations for these values of the parameters, is Abelian.
   In \leref{orbit} we prove that all rational solutions of the Sasano system of type $A^{(2)}_5$ for which $\alpha_0=\alpha_1=\frac{\alpha_3}{2}, \alpha_2=\frac{1}{4}-\alpha_3$
  and $4 \alpha_3-\frac{1}{2} \in \ZZ$ can be obtained from the seed solution \eqref{seed} with $\alpha_3=\frac{1}{8}$ using the B\"acklund transformations \eqref{sym}.
   Then, studying the second normal variational equations along \eqref{seed} for $\alpha_3=\frac{1}{8}$, we obtain the second  key non-integrable result of this paper.

   \bth{key2}
   Assume that $\left(\alpha_0, \alpha_1, \alpha_2, \alpha_3\right)=\left(\frac{1}{16}, \frac{1}{16}, \frac{1}{8}, \frac{1}{8}\right)$.
    Then the Sasano system \eqref{s} is not integrable by rational first integrals.
      \ethe

   The B\"acklund transformations \eqref{sym} are canonical transformations, which are rational in all dynamical variables. With the aim os this fact we can extend
  the results of \thref{key1} and \thref{key2} to the main theorem of this paper.

    \bth{main}
  Assume that the Sasano system of type $A^{(2)}_5$ admits a particular rational solution. Then it is not integrable by rational first integrals.
   \ethe

   This paper is organized as follows. In the next section we briefly recall the basic notations and theorems from the Morales-Ramis-Sim\'o  theory  and Picard-Vessiot theory.
  In section 3 we give a proof of \thref{key1}. In section 4 we provides a proof of \thref{key2}.
  In the last section, using B\"acklund transformations, we generalize the results of the sections 2 and 4 and establish the main result of this paper.
%%%%%%%%%%%%%%%%%%%%%%%%%%%%%%%%%%%%%%%%%%%
%theory
%%%%%%%%%%%%%%%%%%%%%%%%%%%%%%%%%%%%%%%%%%%%%
         \section{Preliminaries}

             \subsection{Non-integrability of Hamiltonian systems and differential Galois theory}

      In this section we briefly recall the Morales-Ramis theory on non-integrability of Hamiltonian systems following \cite{M1, MR, MR1, MRS}.

   Consider a complex analytical Hamiltonian system
          \be\label{HS}
            \dot{x}(t)=X_H(x)
         \ee
        with a Hamiltonian $H : M \rightarrow \CC$ on a complex symplectic manifold $M$ of dimension $2 n$.  Recall that by the theorem of Liouville-Arnold \cite{Ar}
    the Hamiltonian system \eqref{HS} is completely integrable if there exist  $n$  functionally independent first integrals $f_1=H, f_2, f_3, \ldots$ and in involution.
    Let $x(t)$ be a non-equilibrium particular solution of \eqref{HS}.
     The phase curve  $\Gamma$ is the connected Riemann surface  corresponding to this solution. The variational equations $(\textrm{VE})$ along $\Gamma$ have the form
         $$\,
              \dot{\xi}=\frac{\partial X_{H}(x(t))}{\partial x}\,\xi, \quad \xi\in T_{\Gamma} M\,.
       \,$$
         We always can reduce the variational equations using the Hamiltonian in the following sense. Consider the normal bundle of $\Gamma$ on the level variety
    $M_h=\{x\,|\,H(x)=h\}$. The projection of the variational equations on this bundle is called the first normal variational equations $(\textrm{NVE})_1$. The dimension of the $(\textrm{NVE})_1$
 is $2 n -2$.  Assume, in addition, that $x(t)$ is a rational non-equilibrium particular solution of \eqref{HS} and let as above $\Gamma$ be the phase curve corresponding to it.
  Assume also that the differential field $K$ of coefficients of the $(\textrm{NVE})_1$ is the field or rational functions, that is $K=\CC(t)$ and that $t=\infty$ is an irregular singularity for the
 $(\textrm{NVE})_1$. The solutions of the $(\textrm{NVE})_1$ define a Picard-Vessiot extension $L_11$ of the basic  field $K$. This in its turn defines
   a differential Galois group $G_1=\textrm{Gal}(L_1/K)$. Then the main theorem of the Morales-Ramis theory states \cite{M, MR}
        \bth{MR}(\bf{Morales-Ruiz and Ramis})
           If the Hamiltonian system \eqref{HS} is completely integrable by rational first integrals in a neighborhood of $\Gamma$, not necessarily independent on $\Gamma$
              itself, then the connected component $(G_1)^0$ of the unit element of the differential Galois group $G_1$ is Abelian.
         \ethe
  \thref{MR} provides a necessary condition for completely integrability and a sufficient condition for non-integrability of a given Hamiltonian system. That is, if the connected component $(G_1)^0$ of the unit
 element of the Galois group $G_1$ is not Abelian, then the corresponding Hamiltonian system \eqref{HS} is not integrable by rational  first integrals.

     Unfortunately, the opposite is not true in general.
    That is if the connected component $(G_1)^0$ of the unit element of the differential Galois group $G_1$ is
    Abelian, one cannot deduce that the corresponding Hamiltonian system \eqref{HS} is completely integrable. Beyond the first
    variational equations Morales-Ruiz, Ramis and Sim\'{o} suggest in \cite{MRS} to use higher order variational equations
    to solve such integrability problems. Let as above $x(t)$ be a rational non-equilibrium particular solution of the Hamiltonian
    system \eqref{HS}. We write the general solution as $x(t, z)$, where $z$ parametrizes it near $x(t)$ as $x(t, z_0)=x(t)$.
    Then we can write the system \eqref{HS} as
     \be\label{HS1}
      \dot{x}(t, z)=X_H(x(t, z))\,.
     \ee
   Denote by $x^{(k)}(t, z),\,k \geq 1$ the derivatives of $x(t, z)$ with respect to $z$ and by
   $X^{(k)}_H(x),\,k \geq 1$ the derivatives of $X_H(x)$ with respect to $x$. By successive derivations of \eqref{HS1} with respect
   to $z$ and evaluations at $z_0$ we obtain the so called $k$-th variational equations $(\textrm{VE})_k$ along the solution
   $x(t)$
    \be\label{vek}
     \dot{x}^{(k)}(t)=X^{(1)}_H(x(t))\,x^{(k)}(t) + P\left(x^{(1)}(t), x^{(2)}(t), \ldots, x^{(k-1)}(t)\right)\,.
    \ee
   Here $P$ denotes polynomial terms in the monomials of order $|k|$ of the components of its arguments. The coefficients
   of $P$ depend on $t$ through $X^{(j)}_H(x(t)), j < k$. For every $k > 1$ the linear non-homogeneous system \eqref{vek}
   can be arranged as a linear homogeneous system of higher dimension by making the monomials of order $|k|$ in $P$ new
   variables and adding to \eqref{vek} their differential equations. If we restrict the system \eqref{vek} to the variables
   that define the $(\textrm{NVE})_1$ the corresponding linear homogeneous system is the so called $k$-th linearized
   normal variational equations $(\textrm{LNVE})_k$. The solutions of the chain of $(\textrm{LNVE})_k$ define a chain
   of Picard-Vessiot extensions of the main field $K=\CC(t)$ of the coefficients of $(\textrm{NVE})_1$, i.e. we have
   $K \subset L_1 \subset L_2 \subset \cdots \subset L_k$, where $L_1$ is above, $L_2$ is the Picard-Vessiot extension
   of $K$ associated with $(\textrm{LNVE})_2$, etc. Then we can define the differential Galois groups
   $G_1=\textrm{Gal}(L_1/K),\,G_2=\textrm{Gal}(L_2/K), \ldots, G_k=\textrm{Gal}(L_k/K)$. Assume as above that $t=\infty$ is an irregular singularity for the $(\textrm{NVE})_1$
   and therefore for $(\textrm{NVE})_k$ for all $k \geq 2$. Then the main theorem of the
   Morales-Ruiz - Ramis - Sim\'{o} theory states

    \bth{MRS}{\bf (Morales-Ramis-Sim\'{o})}
     If the Hamiltonian system \eqref{HS} is completely integrable with rational first integrals, then for every $k\in\NN$
     the connected component of the unit element $(G_k)^0$ of the differential Galois group $G_k=\textrm{Gal}(L_k/K)$
     is Abelian.
    \ethe

  From \thref{MRS} it follows,  that if we find a group $(G_k)^0$, which is not Abelian, then the Hamiltonian system \eqref{HS} will be non-integrable
   by  rational first integrals.
   Note that this non-commutative group $(G_k)^0$ will be automatically a solvable group. In this way
   non-integrability in the sense of the Hamiltonian dynamics will correspond to integrability in the Picard-Vessiot sense.

       \subsection{The differential Galois group at an irregular singularity}

               In this section we recall the basic definitions and results from the theory of linear differential systems and Picard-Vessiot theory,
     which are required to describe the differential Galois group at an irregular singularity (\cite{MaR, R, Si}).
            All angular directions and sectors are considered on the Riemann surface of the natural logarithm.

      Consider a linear differential system with coefficient rational functions
             \be\label{eq1}
             \eta'(\tau)=A(\tau)\,\eta(\tau), \quad A(\tau)\in M_n(\CC(\tau))\,.
                  \ee
        \bde{gal}
                The differential Galois group $G$ of the system \eqref{eq1} over $\CC(\tau)$ is the group of all differential automorphisms of a Picard-Vessiot extension of
              $\CC(t)$ associated to \eqref{eq1}, which leave fixed the elements of $\CC(\tau)$. This group is isomorphic to an algebraic subgroup of $\textrm{GL}_n(\CC)$
      with respect to a given fundamental matrix solution of \eqref{eq1}
         \ede
           Assume that the system \eqref{eq1} has a non-resonant irregular singularity at $\tau=0$ of Poincar\'e rank 1. The classical theorem of
        Hukuhara-Turrittin-Wasow \cite{W} says that the system \eqref{eq1} possesses a unique formal fundamental matrix solution $\hat{\Phi}(\tau)$ at $\tau=0$
           in the form
                \be\label{ffms}
                      \hat{\Phi}(\tau)=\hat{H}(\tau)\,\tau^{\Lambda}\,\exp \left(\frac{Q}{\tau}\right)\,,
                   \ee
            where
                $$\,
                \Lambda=\diag (\lambda_1, \ldots, \lambda_n), \quad Q=\diag(q_1, \ldots, q_n), \quad  \lambda_j, q_j \in \CC, \,\,1 \leq j \leq n\,.
                   \,$$
  The entries of the matrix $\hat{H}(\tau)$ are in general divergent power series. Consider the system \eqref{eq1} and its formal fundamental matrix solution
       $\hat{\Phi}(\tau)$ over the field $\CC((\tau))$ of formal power series in $\tau$.
               \bde{fm}
                   With respect to the formal fundamental matrix solution $\hat{\Phi}(\tau)$ given by \eqref{ffms}, we define the formal monodromy matrix
          $\hat{M}_0\in \textrm{GL}_n(\CC)$ around the origin as
                      $$\,
                            \hat{\Phi}(\tau . e^{2 \pi\,i})=\hat{\Phi}(\tau)\,\hat{M}_0.
                     \,$$
                   In particular,
                          $$\,
                         \hat{M}_0=e^{2 \pi\,i\,\Lambda}\,.
                           \,$$
              \ede

                    \bde{et}
                    With respect to the formal fundamental matrix solution $\hat{\Phi}(\tau)$ given by \eqref{ffms}, we define the exponential torus $\T$ as the differential Galois group
            $\textrm{Gal}(E/F)$, where $F=\CC((\tau))(\tau^{\lambda_1}, \ldots, \tau^{\lambda_n})$ and $E=F(e^{q_1/\tau}, \ldots, e^{q_n/\tau})$.
                    \ede
                    Since for $\sigma\in \T$ we have that $\sigma(\tau^{\lambda})=\tau^{\lambda}$ and $\sigma (e^{q/\tau})=c\,e^{q/\tau},\,c\in\CC^*$, we can consider $\T$
               as a subgroup of $(\CC^*)^n$.

                  Denote by $\hat{h}_{i j}(\tau)$ the entries of the matrix $\hat{H}(\tau)$, that is $\hat{H}(\tau)=(\hat{h}_{i j}(\tau))_{i, j=1}^n$.
                      \bde{sd}
                         For every divergent power series $\hat{h}_{i j}(\tau)$ we define a family $\Theta_j$ of admissible singular directions
                                    $$\,
                                    \Theta_j=\{\theta_{ji}, 0 \leq \theta_{j i} < 2 \pi\}\,,
                                   \,$$
                 where $\theta_{j i}$ is the bisector of the sector $\left\{\textrm{Re} \left(\frac{q_i-q_j}{\tau}\right) < 0\right\}$. In particular,
                             $$\,
                         \Theta_j=\left\{\theta_{j i}, 0 \leq \theta_{j i} , 2 \pi,\, \theta_{j i}=\arg(q_j - q_i), 1 \leq i \leq n, \,i \neq j\right\}\,.
                            \,$$
                          Denote by $\Theta$ the set of all admissible singular directions of the system \eqref{eq1}, that is
                            $$\,\ds
                                      \Theta =\cup _{j} \Theta _j\,.
                             \,$$
                       \ede

                          Denote $F(\tau)=\tau^{\Lambda}\,\exp \left(\frac{Q}{\tau}\right)$.
                     The applications of summability theory to the ordinary differential equations leads to the following remarkable result
                   \bth{afmt}
                    In the formal fundamental matrix solutions $\hat{\Phi}(\tau)$ from \eqref{ffms} the entries of the matrix $\hat{H}(\tau)$ are 1-summable in every non-singular direction
                      $\theta$. If we denote by $H_{\theta}(\tau)$ the 1-sum of the matrix $\hat{H}(\tau)$ along a nonsingular direction $\theta$ and by
                      $F_{\theta}(\tau)$ the branch of $F(\tau)$ for $\theta=\arg (\tau)$ then the matrix
                $\Phi_{\theta}(\tau)=H_{\theta}(\tau)\,F_{\theta}(\tau)$ gives an actual fundamental matrix solution at the origin of the system  \eqref{eq1} on a small sector bisected by
                 $\theta$.
                  \ethe

              The reader can find needed aspects of the summability theory in works of Loday-Richaud \cite{LR} and Ramis \cite{R}.

                 Let $\epsilon > 0$ be a small number. Let $\theta-\epsilon$ and $\theta+\epsilon$ be two nonsingular neighboring directions of the singular direction $\theta\in \Theta$.
                 Let $\Phi_{\theta-\epsilon}(\tau)$ and $\Phi_{\theta+\epsilon}(\tau)$ be the actual fundamental matrix solutions at the origin of the system \eqref{eq1} related to the
                     directions $\theta-\epsilon$ and $\theta+\epsilon$ in the sense of \thref{afms}. Then

                     \bde{stokes}
                        With respect to the actual fundamental matrix solutions $\Phi_{\theta-\epsilon}(\tau)$ and $\Phi_{\theta+\epsilon}(\tau)$, the Stokes matrix
                        $St_{\theta}\in \textrm{GL}_n(\CC)$ corresponding to the singular direction $\theta$ is defined as
                                 $$\,
                                         St_{\theta}=\left(\Phi_{\theta+\epsilon}(\tau)\right)^{-1}\,\Phi_{\theta-\epsilon}(\tau)\,.
                                \,$$
                        \ede
             The density theorem of Ramis describes the local differential Galois group at $\tau=0$ of the system \eqref{eq1}.

       \bth{R}(\bf{Ramis})
             The differential Galois group at $\tau=0$ of the system \eqref{eq1} is the Zariski closure in $\textrm{GL}_n(\CC)$ of the group generated by the formal monodromy
             $\hat{M}_0$, the exponential torus $\T$, and the Stokes matrices $St_{\theta}$ for all singular directions $\theta$.
     \ethe
 %%%%%%%%%%%%%%%%%%%%%%%%%%%%%%%%%%%%%%%%%
%key theorem1
%%%%%%%%%%%%%%%%%%%%%%%%%%%%%%%%%%%%%%%%%%%%

   \section{ Proof of \thref{key1} }

   In this section we will prove \thref{key1}. When $\alpha_0=\alpha_1=\alpha_3/2, \alpha_2=1/4-\alpha_3$ the Hamiltonian system \eqref{H} becomes
  \be\label{H1}
   \frac{d x}{d s} &=&
   \frac{2\,x^2\,y - x^2  +2 \alpha_3\,x}{t} - 1 + 4\,w +  \frac{2\,x\,z\,w}{t},\nonumber\\[0.15ex]
    \frac{d y}{d s} &=&
    \frac{-2\,x\,y^2 + 2\,x\,y - 2\alpha_3\,y +\alpha_3/2}{t} - \frac{2\,y\,z\,w}{t},\\[0.15ex]
   \frac{d z}{d s} &=&
   \frac{2\,z^2\,w - z^2 +2 \alpha_3\,z}{t} -1 + 4\,y  + \frac{2\,x\,y\,z}{t},\nonumber\\[0.15ex]
    \frac{d w}{d s} &=&
    \frac{-2\, z\, w ^2 + 2 \,z\,w - 2 \alpha_3\,w + \alpha_3/2}{t} - \frac{2\,x \,y\,w}{t},\nonumber\\[0.15ex]
     \frac{d t}{d s} &=& 1,\nonumber\\[0.15ex]
      \frac{d F}{d s} &=&
      \frac{y\,(y-1)\,x^2 + 2 \alpha_3\,x\,y -\alpha_3/2\,x+w\,(w-1)\,z^2 +2 \alpha_3\,z\,w-\alpha_3/2\,z+2 x\,y\,z\,w}{t^2}\,. \nonumber
  \ee
   The system \eqref{H1} possesses the following non-equilibrium particular solution
    \be\label{sol}
               x=z=0,\quad y=w=\frac{1}{4}, \quad t=s, \quad F=0.
      \ee
  Since $t=s$ we use $t$ as an independent variable instead of $s$.

  For the first normal variational equations of the system \eqref{H1} along the solution \eqref{sol} we
   obtain the system
         \be\label{1}
         \dot{x}_1   &=&
                \frac{2 \alpha_3}{t}\,x_1  + 4 w_1,\nonumber\\[0.15ex]
        \dot{y}_1  &=&
    \frac{3}{8 t}\,x_1 - \frac{2 \alpha_3}{t}\,y_1  - \frac{1}{8 t}\,z_1,\\[0.15ex]
         \dot{z}_1   &=&
               4 y_1    + \frac{2 \alpha_3}{t}\,z_1,\nonumber\\[0.15ex]
        \dot{w}_1   &=&
         - \frac{1}{8 t}\,x_1   + \frac{3}{8 t}\,z_1 - \frac{2 \alpha_3}{t}\,w_1.\nonumber
       \ee
   The transformation $t=\tau^2$ takes the system \eqref{1} into the system
        \be\label{nve}
            \tau^{-1}\,X'(\tau)=A(\tau)\,X(\tau),\quad '=\frac{d}{d \tau}
                          \ee
  where $X(\tau)=(x_1(\tau), y_1(\tau), z_1(\tau), w_1(\tau))^T$ and
             $$\ds
            A(\tau)=\left(\begin{array}{cccc}
               \frac{4 \alpha_3}{\tau^2}           & 0             & 0             & 8\\[0.25ex]
               \frac{3}{4 \tau^2}                        & - \frac{4 \alpha_3}{\tau^2}      &-\frac{1}{4 \tau^2}      & 0 \\[0.25ex]
               0                                                     & 8            & \frac{4 \alpha_3}{\tau^2}                                & 0\\[0.25ex]
               -\frac{1}{4 \tau^2}                       & 0            & \frac{3}{4 \tau^2}                                             & - \frac{4 \alpha_3}{\tau^2}
                                     \end{array}\right).
            $$
      We will study the differential Galois group of the system \eqref{nve} instead of the system \eqref{1}. The system \eqref{nve} has two singular points
  over $\CC\PP^1$ : $\tau=0$ and $\tau=\infty$. The point $\tau=0$ is a regular singularity and the point $\tau=\infty$ is an irregular singularity of Poincar\'{e} rank 1.

        \bth{1}
           The transformation
                   \be\label{tr}
                         X(\tau)=T_1\,S_1(\tau)\,T_2\,S_2(\eta)\,T_3\,V(\eta),\quad \tau=\frac{1}{4}\,\eta
                       \ee
       reduces the system \eqref{nve} to the system
                       \be\label{sys}\ds
                      \eta^{-1}\,\frac{d V(\eta)}{d \eta}=\left(\begin{array}{cccc}
                              \frac{3}{\eta^2} -1    & \frac{8 \alpha_3-1}{\eta^2}                   & 0                  & 0\\[0.25ex]
                               \frac{8 \alpha_3-1}{\eta^2}     &\frac{3}{\eta^2} +1                 & 0                  & 0\\[0.25ex]
                                        0                                         & 0                                               & \frac{3}{\eta^2} - i\,\sqrt{2}      & \frac{8 \alpha_3-1}{\eta^2}\\[0.25ex]
                                        0                                         & 0                                               & \frac{8 \alpha_3-1}{\eta^2}       & \frac{3}{\eta^2} + i\,\sqrt{2}
                               \end{array}\right)\,V(\eta)\,.
                          \ee
 The matrices $S_1(\tau)$ and $S_2(\eta)$ are diagonal matrices given by
                           $$\,
                           S_1(\tau)=\diag (1, \tau^{-1/2}, \tau^{-1}, \tau^{-3/2}),\quad
                      S_2(\eta)=\diag (1, \eta^{-1}, \eta^{-2}, \eta^{-3})\,.
                                          \,$$
       The matrices $T_j, j=1, 2, 3$ are constant invertible matrices given by
         \ben
      T_1=\left(\begin{array}{cccc}
                             8   & 0   & 0   & 0\\
                             0   & 0   & 0   & 1\\
                             0   & 0   & 8   & 0\\
                             0   & 1   & 0   & 0
                              \end{array}\right), \quad
                 T_2=\left(\begin{array}{cccc}
                             0   & 0   & 2/3   & 0\\
                             0   & 0   & 0   & 8/3\\
                             1/4   & 0   & 0   & 0\\
                             0   & 1   & 0   & 0
                              \end{array}\right),\\[0.25ex]
               T_3=\left(\begin{array}{rcrr}
                             1/4    & 1/4   & 1/4   & 1/4\\
                             -1/4   & 1/4   & -i\,\sqrt{2}/4   & i\,\sqrt{2}/4\\
                             3/8   & 3/8   & -3/8   & -3/8\\
                             -3/8   & 3/8   & 3 i\,\sqrt{2}/8   & -3 i\,\sqrt{2}/8
                              \end{array}\right).
                                 \een
             \ethe

          \proof
       The proof is straightforward.\qed

            \bre{tr}
             The transformation \eqref{tr} is a chain of shearing (by the matrices $S_1(\tau)$ and $S_2(\eta)$) and similar ( by the matrices $T_j, j=1, 2, 3$) transformations.
              \ere
      \bth{G}
         Assume that $4 \alpha_3 - \frac{1}{2}\in\ZZ$. Then the connected component  $(G_1)^0$ of the unit element of the differential Galois group $G_1$
     of the system \eqref{nve} is Abelian. Otherwise $(G_1)^0=SL_2(\CC) \times SL_2(\CC)$, that is $(G_1)^0$ is not Abelian.
     \ethe

   \proof
  Denote $V(\eta)=(v_1(\eta), v_2(\eta), v_3(\eta), v_4(\eta))^T$. When $8 \alpha_3-1 \neq 0$ the system \eqref{sys} is equivalent to the following independent
  second order scalar equations
                  \be\label{A}
                       & &
           v_1'' - \frac{5}{\eta}\,v_1' + \left[2 - \eta^2 - \frac{(8 \alpha_3-1)^2 -9}{\eta^2}\right]\,v_1=0,\\[0.25ex]\label{B}
                       & &
           v_3'' - \frac{5}{\eta}\,v_3' + \left[2 i\,\sqrt{2} + 2 \eta^2 - \frac{(8 \alpha_3-1)^2 -9}{\eta^2}\right]\,v_3=0.
                    \ee
         The change $\eta^2=\xi$ takes the equations \eqref{A} and \eqref{B} into the equations
        \be\label{A1}
                       & &
           \frac{d^2 v_1}{d \xi^2} - \frac{2}{\xi}\,\frac{d v_1}{d \xi} + \left[-\frac{1}{4} +\frac{1}{2 \xi} - \frac{(8 \alpha_3-1)^2 -9}{4 \xi^2}\right]\,v_1=0,\\[0.25ex]\label{B1}
                       & &
           \frac{d^2 v_3}{d \xi^2} - \frac{2}{\xi}\,\frac{d v_3}{d \xi} + \left[\frac{1}{2} + \frac{ i\,\sqrt{2}}{2 \xi}  - \frac{(8 \alpha_3-1)^2 -9}{4 \xi^2}\right]\,v_3=0.
           \ee
               Next, the transformations $v_j(\xi)=\xi\,\omega_j(\xi), j=1, 3$ take the equations \eqref{A1} and \eqref{B1} into their normal forms
               \be\label{A2}
                       & &
           \frac{d^2 \omega_1}{d \xi^2}  + \left[-\frac{1}{4} +\frac{1}{2 \xi} + \frac{1/4 - \left((8 \alpha_3-1)/2\right)^2 }{ \xi^2}\right]\,\omega_1=0,\\[0.25ex]\label{B2}
                       & &
           \frac{d^2 \omega_3}{d \xi^2}  + \left[\frac{1}{2} + \frac{ i\,\sqrt{2}}{2 \xi}  + \frac{1/4 -\left((8 \alpha_3-1)/2\right)^2}{ \xi^2}\right]\,\omega_3=0.
           \ee
 The equation \eqref{A2} is the Whittaker equation \cite{WW}
                    $$\,
    \frac{d^2 \omega}{d \xi^2} + \left[-\frac{1}{4} + \frac{\kappa}{\xi} + \frac{1/4 - \mu^2}{\xi^2}\right]\,\omega=0
                       $$
 with $\kappa=1/2$ and $\mu=(8 \alpha_3-1)/2$. The equation \eqref{B2} is not a Whittaker equation but the change $\xi=-\frac{i}{\sqrt{2}}\,\zeta$ takes it into
the Whittaker equation with the same parameters $\kappa=1/2,\,\mu=(8 \alpha_3-1)/2$. The differential Galois group of the Whittaker equation is well determined by
 Martinet and Ramis in \cite{MaR}. They show that the differential Galois group of the Whittakes equation is in fact determined by the local differential Galois group at
the irregular singularity at $\xi=\infty$. Following \cite{MaR} we can construct a local formal fundamental set of solutions $\{\hat{\omega}_{11}(\xi), \hat{\omega}_{12}(\xi)\}$
at $\xi=\infty$ of the equation \eqref{A2}
  \be\label{fss}
    \hat{\omega}_{11}(\xi)=\xi^{1/2}\,e^{-\xi/2}\,\hat{\phi}_1(\xi),\qquad
   \hat{\omega}_{12}(\xi)=\xi^{-1/2}\,e^{\xi/2}\,\hat{\phi}_2(\xi),
  \ee
        where
  \be\label{ps}
       \hat{\phi}_1(\xi)  &=& 1 + \sum_{n=1}^{\infty} \frac{\left(4 \alpha_3 - \frac{1}{2}\right)^2\,\left[\left(4 \alpha_3-\frac{1}{2}\right)^2 - 1^2\right] \ldots
         \left[\left(4 \alpha_3 - \frac{1}{2}\right)^2 - (n-1)^2\right]}{n!}\,\xi^{-n},\nonumber\\[0.3ex]
       \hat{\phi}_2 (\xi) &=& 1 + \sum_{n=1}^{\infty} \frac{\left[\left(4 \alpha_3-\frac{1}{2}\right)^2 - 1^2\right] \ldots
         \left[\left(4 \alpha_3 - \frac{1}{2}\right)^2 - n^2\right]}{n!}\,(-\xi)^{-n}.
 \ee
   Obviously when $4 \alpha_3-1/2\in\NN$ or $4 \alpha_3-1/2\in -\NN$ the power series $\hat{\phi}_1(\xi)$ and $\hat{\phi}_2(\xi)$ are together polynomials in $\xi^{-1}$. In this case the local
differential Galois group of the Whittaker equation at $\xi=\infty$, and therefore  its whole differential Galois group $G$, is Abelian \cite{MaR}. In particular, in this case
  $$\,
      G^0=\left\{\left(\begin{array}{cc}
                \lambda            & 0\\
                   0                   & \lambda^{-1}
                    \end{array}\right), \quad \lambda\in\CC^* \right\}.
  \,$$
          When $4 \alpha_1-1/2\notin \NN$ or $4\alpha_3-1/2 \notin - \NN$ both power series $\hat{\phi}_1(\xi)$ and $\hat{\phi}_2(\xi)$ are divergent and we observe non-trivial Stokes phenomena.
  In this case the differential Galois group of the equation \eqref{A2} is the group $SL_2(\CC)$.

  The transformations $\eta^2=\xi, \,\,v_j(\xi)=\xi\,\omega_j(\xi), j=1, 3$ and $\xi=-\frac{i}{\sqrt{2}}\,\zeta$ (for the equation \eqref{B2}) change the differential Galois groups  of the equations
  \eqref{A} and \eqref{B} but they preserve the  corresponding connected components of their unit elements. Since the scalar equations \eqref{A} and \eqref{B} and the corresponding linear systems
 including in the system \eqref{sys} belong to the same differential modules their differential Galois groups coincide \cite{Si}.

   We complete the proof considering the case when $8 \alpha_3 -1 =0$ or equivalently $4 \alpha_3-1/2=0$. In this save the system \eqref{sys} splits into 4 solvable first order scalar equations
   \ben
   v_1'=\left(\frac{3}{\eta}-\eta\right)\,v_1,  & &       v_2'=\left(\frac{3}{\eta}+\eta\right)\,v_2,\\[0.3ex]
     v_3'=\left(\frac{3}{\eta}-i\,\sqrt{2}\,\eta\right)\,v_3,  & &    v_4'=\left(\frac{3}{\eta}+i\,\sqrt{2}\,\eta\right)\,v_4.
   \een
As a result we have that  when $4 \alpha_3-1/2\in\ZZ$ the connected component $(G_1)^0$ of the unit element of the differential Galois group of the system \eqref{sys} is Abelian and in particular
   $$\,
    (G_1)^0=\left\{\left(\begin{array}{cccc}
                     \lambda_1            & 0    & 0            & 0\\
                           0                       &\lambda_1^{-1}          & 0              & 0\\
                           0                       & 0                                    & \lambda_2               & 0\\
                           0                       & 0                                    & 0                                 & \lambda_2^{-1}
              \end{array}\right),\quad \lambda_1, \lambda_2\in\CC^*, \,\,\lambda_1, \lambda_2 \quad \textrm{are not roots of the unit}\right\}.
   \,$$
      When $4 \alpha_3-1/2\notin\ZZ$ the group $(G_1)^0$ is not Abelian and in particular, $(G_1)^0=SL_2(\CC) \times SL_2(\CC)$.

    Finally, the proof follows from the fact that  the transformation \eqref{tr} preserves the connected component of the unit element of the differential Galois group of the system \eqref{nve}.

 \qed

 {\it Proof of \thref{key1}}. The proof follows from \thref{G} and he observation that the transformation $t=\tau^2$ preserves the connected component of the unit element of the
  differential Galois group of first normal variational equations \eqref{1}. \qed

%%%%%%%%%%%%%%%%%%%%%%%%%%%%%%%%%%%%%%%%%%
%key theorem2
%%%%%%%%%%%%%%%%%%%%%%%%%%%%%%%%%%%%%%%%%%%%

    \section{ Proof of \thref{key2}}

        In this section we will prove the key \thref{key2} by studying the second normal variational equations of the Hamiltonian system \eqref{H1} along the solution \eqref{sol}.

   For the second normal variational equations we obtain the following fourth-order system of non-homogeneous equations
     \ben
             \dot{x}_2     &=&  \frac{2 \alpha_3}{t}\,x_2 + 4\,w_2 - \frac{1}{2 t}\,x_1^2 + \frac{1}{2 t}\,x_1\,z_1,\\[0.15ex]
             \dot{y}_2     &=& \frac{3}{8 t}\,x_2 - \frac{2 \alpha_3}{t}\,y_2 - \frac{1}{8 t}\,z_2 + \frac{1}{t}\,x_1\,y_1 - \frac{1}{2 t}\,y_1\,z_1 - \frac{1}{2 t}\,w_1\,z_1,\\[0.15ex]
              \dot{z}_2    &=& 4 \,y_2 + \frac{2 \alpha_3}{t}\,z_2 - \frac{1}{2 t}\,z_1^2 + \frac{1}{2 t}\,x_1\,z_1,\\[0.15ex]
            \dot{w}_2    &=& -\frac{1}{8 t}\,x_2 + \frac{3}{8 t}\,z_2 - \frac{2 \alpha_3}{t}\,w_2 - \frac{1}{2 t}\,x_1\,y_1 - \frac{1}{2 t}\,x_1\,w_1 + \frac{1}{t}\,z_1\,w_1\,.
      \een
 This system can be arranged as a homogeneous linear system of fourteenth-order. Instead of this higher-order system we will study a fifth-order linear homogeneous system obtained from
 the $(\textrm{NVE})_2$ in the following way. We put
  \be\label{ch}
   q:=x_2+z_2,\quad p:=y_2+w_2,\quad u:=(x_1-z_1)\,(y_1-w_1),\quad v:=(x_1-z_1)^2, \\[0.15ex]
            g:=(y_1-w_1)^2, \quad \tau^2:=1/t\,.
  \ee
 Then the $(\textrm{LNVE})_2$ are equivalent to the following fifth-order linear homogeneous system
           \be\label{s5}
          q'  &=& -\frac{4 \alpha_3}{\tau}\,q - \frac{8}{\tau^3} \,p + \frac{1}{\tau}\,v,\nonumber\\[0.15ex]
          p'  &=& -\frac{1}{2 \tau}\,q +\frac{4 \alpha_3}{\tau}\,p - \frac{1}{\tau}\,u,\\[0.15ex]
           v'  &=& -\frac{8 \alpha_3}{\tau}\,v + \frac{16}{\tau^3}\,u\nonumber\\[0.15ex]
        u'  &=& -\frac{1}{\tau} \,v + \frac{8}{\tau^3}\,g,\nonumber\\[0.15ex]
          g'  &=& - \frac{2}{\tau}\,u + \frac{8 \alpha_3}{\tau}\,g\,,\nonumber
            \ee
      where $'=\frac{d}{d \tau}$. The system \eqref{s5} has two singularities over $\CC\PP^1$ : the point $\tau=0$ is an irregular singularity of
     Poincar\'e rank 1, while the point $\tau=\infty$ is a regular singularity. In what follows we study the local differential Galois group at the origin
     of the system \eqref{s5}.

    From here on we assume that $4 \alpha_3=n + \frac{1}{2}$, where $n\in\ZZ$.
    The system \eqref{s5} is equivalent to the following second-order non-homogeneous linear differential equation
              \be\label{E}
                            q'' + \frac{3}{\tau}\,q' - \left[\frac{4}{\tau^4} + \frac{(2 n+1)\,(2 n-3)}{4 \tau^2}\right]\,q=
                                     \frac{3}{2 \tau}\,v' + \frac{2}{\tau^2}\,v\,.
                 \ee
                If we find 3 particular solutions of the equation \eqref{E} then we will build a fundamental matrix solution of the system \eqref{s5}.

  The variables $x_1-z_1$ and $x_1+z_1$  satisfy the following second-order linear homogeneous differential equations
     \be\label{minus}
              & &
   (x_1-z_1)'' + \frac{3}{\tau}\,(x_1-z_1)' + \left[\frac{8}{\tau^4} - \frac{(2 n+1)\,(2 n-3)}{4 \tau^2}\right]\,(x_1-z_1)=0\,,\\[0.35ex]
          & &
             (x_1+z_1)'' + \frac{3}{\tau}\,(x_1+z_1)' - \left[\frac{4}{\tau^4} - \frac{(2 n+1)\,(2 n-3)}{4 \tau^2}\right]\,(x_1+z_1)=0\,\label{plus}
    \ee
     respectively.

  We have

   \ble{l1}
  The equations \eqref{minus} and \eqref{plus}  have local fundamental sets  of solutions at $\tau=0$ in the form
     \be\label{l1}
 \left\{\tau^{-1/2}\,e^{-2\,i\,\sqrt{2}/\tau}\,f_1,\,\tau^{-1/2}\,e^{2\,i\,\sqrt{2}/\tau}\,f_2\right\} \quad \textrm{and} \quad
      \left\{\tau^{-1/2}\,e^{-2/\tau}\,g_1,\,\tau^{-1/2}\,e^{2/\tau}\,g_2\right\}\,,
 \ee
    respectively.
   Depending  on $n$ the polynomials $f_j, g_j, \,j=1,2 $ are given by
              \begin{enumerate}
                       \item\,
                    If $n \in \NN$ then
           \ben
                   & &
               f_1=\sum_{k=0}^{n-1} a_k\,\tau^k,\quad
                     f_2=\sum_{k=0}^{n-1} (-1)^k\,a_k\,\tau^k\,,\\[0.15ex]
                    & &
               g_1=\sum_{k=0}^{n-1} b_k\,\tau^k,\quad
                     g_2=\sum_{k=0}^{n-1} (-1)^k\,b_k\,\tau^k\,.
              \een

                      \item
                    If $n\in\ZZ_{\leq 0}$ then
            \ben
                      & &
               f_1=\sum_{k=0}^{-n} a_k\,\tau^k,\quad
                     f_2=\sum_{k=0}^{-n} (-1)^k a_k\,\tau^k\,,\\[0.15ex]
                    & &
               g_1=\sum_{k=0}^{-n} b_k\,\tau^k,\quad
                     g_2=\sum_{k=0}^{-n} (-1)^k\,b_k\,\tau^k\,.
              \een

        \end{enumerate}
      where
            \ben
                    a_k= (-1)^k\,\frac{n^{(k)}\,(-n+1)^{(k)}}{(4\,i\,\sqrt{2})^k\,k!}, \quad
                         b_k= (-1)^k\,\frac{n^{(k)}\,(-n+1)^{(k)}}{4^k\,k!}\,.
             \een
                       \ele

              \proof
         The change $\tau=-\frac{4\,i\,\sqrt{2}}{\xi},\,x_1-z_1=\xi^{1/2}\,u$ takes the equation \eqref{minus} into the Whittaker equation
           $$
               \frac{d^2 u}{d \xi^2} + \left[-\frac{1}{4} - \frac{n\,(n-1)}{\xi^2}\right]\,u=0.
            $$
     It is well known \cite{MaR} that such an Whittaker equation has a local fundamental set of solutions at $\xi=\infty$ in the form
            $$
               u_1(\xi)=e^{-\xi/2}\,f_1(\xi), \quad u_2(\xi)=e^{\xi/2}\,f_2(\xi)\,.
           $$
          Then the equation \eqref{minus} will have a local fundamental set of solutions at $\tau=0$ in the form \eqref{l1}.

          Looking for solutions $x_1-z_1=\tau^{-1/2}\,e^{-2\,i\,\sqrt{2}/\tau}\,f_1$ and  $x_1-z_1=\tau^{-1/2}\,e^{2\,i\,\sqrt{2}/\tau}\,f_2$ of the equation
         \eqref{minus}, we find that $f_1$ and $f_2$ must satisfy the equations
            \be\label{f12}
                        & &
               \tau^2\,f_1'' + \left[4\,i\,\sqrt{2} + 2\,\tau\right]\,f_1' - n\,(n-1)\,f_1=0,\\[0.15ex]
                        & &
                 \tau^2\,f_2'' + \left[-4\,i\,\sqrt{2} + 2\,\tau\right]\,f_2' - n\,(n-1)\,f_2=0,\nonumber
           \ee
 respectively. Now, it is not difficult to show that $f_1$ and $f_2$ have the pointed forms. In the same manner one can derive a local fundamental set of solutions at
 $\tau=0$ of the equation  \eqref{plus}.
          \qed

               \ble{l2}
                     The polynomial $f=f_1\,f_2$ has the form
                              $$\,
                                        f=\sum_{k=0}^{n-1}  c_{2 k}\,\tau^{2 k}, \quad \textrm{when} \quad n\in\NN
                           \,$$
      or
                              $$\,
                                        f=\sum_{k=0}^{-n}  c_{2 k}\,\tau^{2 k}, \quad \textrm{when} \quad n\in\ZZ_{\leq 0},
                           \,$$
               where
                              $$\,
                                              c_{2 k}=(-1)^k\, \frac{(2 k-1)!!\,n^{(k)}\,(-n+1)^{(k)}}{4^{2 k}\,k!},\quad c_0=1\,.
                                  \,$$
           \ele

            \proof
                From $f_1=\sum_{k=0}^{\sigma} a_k\,\tau^k,\,\,f_2=\sum_{k=0}^{\sigma} (-1)^k\,a_k\,\tau^k,\,\sigma\in\{n-1, -n\}$ it follows that
                        $$\,
                             f=f_1\,f_2=\sum_{k=0}^{\sigma} c_{2k}\,\tau^{2 k}\,.
                             \,$$
         Using the differential equations \eqref{f12} for the polynomials $f_j, j=1, 2$, we find that the polynomial $f$ satisfies the following third-order ordinary differential equation
                  \be\label{f}
                \tau^4\,f''' + 6 \tau^3\,f'' - \left[2 (2 n^2 - 2 n -3)\,\tau^2 -32\right]\,f' - 4 n\,(n-1)\,\tau\,f=0\,.
                  \ee
                 Looking for $f$ in the above form, we find that the coefficients $c_{2 k}$ must satisfy the relation
                         $$\,
                                               c_{2 k}=\frac{(2 k-1)\,(n-k)\,(n+k-1)}{4^2\,k}\,c_{2 k-2}\,.
                                \,$$
                 Then choosing $c_0=1$, we prove the lemma.
  \qed

            Using the fundamental set of solutions, given by \leref{l1}, we choose the following basis for the variable $v=(x_1-z_1)^2$
            \be\label{v}
            v_1=\tau^{-1}\,f,\quad v_2=\tau^{-1}\,e^{-4\,i\,\sqrt{2}/\tau}\,f_1^2, \quad
                  v_3=\tau^{-1}\,e^{4\,i\,\sqrt{2}/\tau}\,f_2^2\,.
             \ee
       With this basis we associate three right hand sides $F=\frac{3}{2 \tau}\,v' + \frac{2}{\tau^2}\,v$ of the equation \eqref{E}
                 \ben
            F_1 &=&
 \frac{1}{2 \tau^3}\,f + \frac{3}{2 \tau^2}\,f',\\[0.8ex]
           F_2 &=&
     \left[ \frac{1}{2 \tau^3} + \frac{6 i\,\sqrt{2}}{\tau^4}\right]\,e^{-4\,i\,\sqrt{2}/\tau}\,f_1^2 + \frac{3}{2 \tau^2}\,e^{-4\,i\,\sqrt{2}/\tau}\,(f_1^2)',\\[0.8ex]
             F_3  &=&
                \left[ \frac{11}{2 \tau^3} - \frac{6 i\,\sqrt{2}}{\tau^4}\right]\,e^{4\,i\,\sqrt{2}/\tau}\,f_2^2 + \frac{3}{2 \tau^2}\,e^{4\,i\,\sqrt{2}/\tau}\,(f_2^2)'\,.
                           \een

                  The next lemma allows us to reduce the problem of integrability of the Sasano system \eqref{s} for
          $$\,
          (\alpha_0, \alpha_1, \alpha_2, \alpha_3)=\left(\frac{2 n+1}{16}, \frac{2 n+1}{16}, \frac{1-2n}{8}, \frac{2 n+1}{8}\right)
          \,$$
      to the same problem but for
                    $$\,
          (\alpha_0, \alpha_1, \alpha_2, \alpha_3)=\left(\frac{1}{16}, \frac{1}{16}, \frac{1}{8}, \frac{1}{8}\right)\,.
          \,$$

                    \ble{orbit}
                     Let
             $$
               (x, y, z, w; \alpha_0, \alpha_1, \alpha_2, \alpha_3)=\left(0, \frac{1}{4}, 0, \frac{1}{4}; \frac{1}{16}, \frac{1}{16}, \frac{1}{8}, \frac{1}{8}\right)
            $$
                  be a rational solution of the Sasano system \eqref{s}. Then starting with it, using the B\"acklund transformations \eqref{sym}, we obtain again a rational solution
        of the system \eqref{s}, for which $\alpha_3=\frac{2 n+1}{8}, n\in\ZZ$ and the parameters $\alpha_j, j=0, 1, 2, 3$ satisfy one of the first four conditions of \thref{M}.
                        \ele

                \proof

   Following Matsuda \cite{Mat}, we define the shift operators $T_j, j=0, 1, 2$ by
         $$\,
                        T_0=\pi\,s_2\,s_3\,s_2\,s_1\,s_0, \qquad  T_1=s_0\,T_0\,s_0, \qquad T_2=s_2\,T_0\,s_2,
              \,$$
 respectively. They act on the parameters $\alpha_j, j=0, 1, 2, 3$ as
        \ben
              T_0(\alpha_0, \alpha_1, \alpha_2, \alpha_3)   &=&
                   (\alpha_0+1/2, \alpha_1+1/2, \alpha_2-1/2, \alpha_3),\\[0.15ex]
              T_1(\alpha_0, \alpha_1, \alpha_2, \alpha_3)   &=&
                   (\alpha_0-1/2, \alpha_1+1/2, \alpha_2, \alpha_3),\\[0.15ex]
             T_2(\alpha_0, \alpha_1, \alpha_2, \alpha_3)   &=&
                   (\alpha_0, \alpha_1, \alpha_2+1/2, \alpha_3-1)\,.
             \een
              We firstly will show how from $\alpha_3=1/8$ we can obtain $\alpha_3=3/8,\,\alpha_3=5/8$ and $\alpha_3=7/8$ using B\"acklund transformation \eqref{sym} and
         the shift operators $T_j, j=0, 1, 2$. Denote by $\alpha^0=\left(\frac{1}{16}, \frac{1}{16}, \frac{1}{8}, \frac{1}{8}\right)$ the vector of the initial values of the parameters $\alpha_j, j=0, 1,  2, 3$.
       Note that the vector $\alpha^0$ satisfies the first condition of \thref{M}. Under the transformation $s_2$ the vector $\alpha^0$ becomes
                 $\alpha=\left(\frac{3}{16}, \frac{3}{16}, -\frac{1}{8}, \frac{3}{8}\right)$, which also satisfies the first condition of \thref{M}.
  Under the transformation $T_2\,s_3$ the vector $\alpha^0$ consecutively becomes
                    $$\,
                \alpha^0  \longrightarrow \alpha^1=\left(\frac{1}{16}, \frac{1}{16}, \frac{5}{8}, -\frac{7}{8}\right) \longrightarrow
                 \alpha^2=\left(\frac{1}{16}, \frac{1}{16}, -\frac{1}{4}, \frac{7}{8}\right)\,.
                    \,$$
         The vector $\alpha^1$ satisfies the first condition, while the vector $\alpha^2$ satisfies the fourth condition of \thref{M}. Similarly, the action of the transformation $T_2\,s_3$ on
 the vector $\alpha$ consecutively leads to
                                   $$\,
                \alpha  \longrightarrow \alpha^4=\left(\frac{3}{16}, \frac{3}{16}, \frac{3}{8}, -\frac{5}{8}\right) \longrightarrow
                 \alpha^5=\left(\frac{3}{16}, \frac{3}{16}, -\frac{1}{4}, \frac{5}{8}\right)\,.
                    \,$$
               Now, applying consecutively the transformations $s_3, s_3\,T_2$ and $s_3\,T_2\,s_3$ to the vectors $\alpha, \alpha^j, j=0,1, 2, 3, 4, 5$ we obtain all values of the parameters, for which
        $\alpha_3=\frac{2 n+1}{8}, n\in\ZZ$.
        \qed

So, to the end of this section
    we assume, that $\alpha_3=\frac{1}{8}$ (resp. $n=0$).
                  \bpr{p1}
                          Assume that  $n=0$. Then depending on $F$ the equation \eqref{E} possesses particular solutions $q_j(\tau), j=1, 2, 3$ in the form
                             \ben
                               q_1(\tau) &=& -\frac{1}{16}\,[ \hat{\phi}_1(\tau) + \hat{\phi}_2(\tau)],\\[0.2ex]
                               q_2(\tau)  &=&
               -\frac{i\,\sqrt{2}}{6}\,\exp\left(\frac{-4\,i\,\sqrt{2}}{\tau}\right) + \exp\left(-\frac{4\,i\,\sqrt{2}}{\tau}\right)\,\left[\frac{5+i\,\sqrt{2}}{432}\hat{\varphi}_1(\tau) +
                       \frac{5-i\,\sqrt{2}}{432}\hat{\varphi}_2(\tau)\right],\\[0.2ex]
                                q_3(\tau) &=&
                                \frac{i\,\sqrt{2}}{6}\,\exp\left(\frac{4\,i\,\sqrt{2}}{\tau}\right) + \exp\left(\frac{4\,i\,\sqrt{2}}{\tau}\right)\,
           \left[\frac{5-i\,\sqrt{2}}{432}\hat{\psi}_1(\tau) + \frac{5+i\,\sqrt{2}}{432}\hat{\psi}_2(\tau)\right],
                         \een
                         where
                     \ben
                      \hat{\phi}_1(\tau) &=&
               \sum_{k=0}^{\infty} \frac{(2 k+1)!!}{4^k}\,\tau^{k+1}, \quad
                     \hat{\phi}_2(\tau)= \sum_{k=0}^{\infty} (-1)^k\,\frac{(2 k+1)!!}{4^k}\,\tau^{k+1}, \\[0.5ex]
                      \hat{\varphi}_1(\tau) &=&
                        \sum_{k=0}^{\infty} \frac{(2 k+1)!!}{2^k\,(2-4\,i\,\sqrt{2})^k}\,\tau^{k+1} ,\quad
                      \hat{\varphi}_2(\tau) =
                                 \sum_{k=0}^{\infty} (-1)^k\,\frac{(2 k+1)!!}{2^k\,(2+4\,i\,\sqrt{2})^k}\,\tau^{k+1} ,\\[0.5ex]
                    \hat{\psi}_1(\tau) &=&
                        \sum_{k=0}^{\infty} \frac{(2 k+1)!!}{2^k\,(2+4\,i\,\sqrt{2})^k}\,\tau^{k+1},\quad
                     \hat{\psi}_2(\tau)=
                                 \sum_{k=0}^{\infty} (-1)^k\,\frac{(2 k+1)!!}{2^k\,(2-4\,i\,\sqrt{2})^k}\,\tau^{k+1}\,.
                        \een
                   \epr

                \proof

                  Consider the equation
                               $$\,
                           \tau^4\, q'' + 3\,\tau^3\,q' - \left[4 - \frac{3}{4}\, \tau^2\right]\,q=
                                    \frac{\tau}{2},
                               \,$$
    which is the equation \eqref{E} with $n=0$  and $F=F_1=\frac{1}{2 \tau^3}$. This equation admits a particular power series solution $q_1(\tau)$ in the form
                           \ben
            q_1(\tau)  &=&
              -\frac{1}{8}\left[\tau + \frac{3.5}{4^2}\,\tau^3 + \frac{3.5.7.9}{4^4}\,\tau^5 + \cdots \right]=-\frac{1}{8} \sum_{k=0}^{\infty}
                        \frac{3.5.7\ldots (4 k+1)}{4^{2 k}}\,\tau^{2 k+1}=\\[0.3ex]
                            &=&
                     -\frac{1}{16} \sum_{k=0}^{\infty} \frac{(2 k+1)!!}{4^k}\,\tau^{k+1}
                    -\frac{1}{16} \sum_{k=0}^{\infty} (-1)^k\,\frac{(2 k+1)!!}{4^k}\,\tau^{k+1}.
                              \een
                  Consider the equation
                               $$\,
                            q'' + \frac{3}{\tau}\,q' - \left[\frac{4}{\tau^4} - \frac{3}{4 \tau^2}\right]\,q=
                                    \left[ \frac{1}{2 \tau^3} + \frac{6 i\,\sqrt{2}}{\tau^4}\right]\,e^{-4\,i\,\sqrt{2}/\tau}\,.
                               \,$$
           Looking for a solution $q_2(\tau)=e^{-4\,i\,\sqrt{2}/\tau}\,h(\tau)$, we find that $g(\tau)$ satisfies the equation
                  $$
               \tau^4\,h''(\tau) + \left[8 i\,\sqrt{2}\,\tau^2 + 3 \tau^3\right]\,h'(\tau) +
                     \left[-36 + 4 i\,\sqrt{2}\,\tau + \frac{3 \tau^2}{4}\right]\,h(\tau)=\frac{\tau}{2} + 6 i\,\sqrt{2}.
                  $$
          This equation has a power series solution
          $$
              h(\tau)=a_0 +\sum_{k=1}^{\infty} a_k\,\tau^k=a_0 + \hat{\varphi}(\tau)\,,
            $$
             where the coefficients $a_k$ satisfy the relation
                       \be\label{cf}
                               & &
                \frac{(2 k-1)\,(2 k-3)}{4}\,a_{k-2} + 4\,i\,\sqrt{2}\,(2 k-1)\,a_{k-1} - 36\,a_k=0,\quad  k \geq 2,\\[0.15ex]
                                 & &
                          a_0=-\frac{i\,\sqrt{2}}{6}, \quad a_1=\frac{5}{6.36},\quad a_2=\frac{11\,i\,\sqrt{2}}{2. 36^2}\,. \nonumber
                        \ee
                It is not so difficult to show that the power series $\hat{\varphi}(\tau)$ has the form
                   \ben
                  \hat{\varphi}(\tau)  &=&
            \frac{5+i\,\sqrt{2}}{432}\,\left[\tau + \frac{3}{2}\,\frac{\tau^2}{2-4\,i\,\sqrt{2}} + \cdots + \frac{(2 k+1)!!}{2^k}\,\frac{\tau^{k+1}}{(2-4\,i\,\sqrt{2})^k} + \cdots\right] + \\[0.2ex]
                                           &+&
                  \frac{5-i\,\sqrt{2}}{432}\,\left[\tau - \frac{3}{2}\,\frac{\tau^2}{2+4\,i\,\sqrt{2}} + \cdots + (-1)^k\,\frac{(2 k+1)!!}{2^k}\,\frac{\tau^{k+1}}{(2+4\,i\,\sqrt{2})^k} + \cdots\right]\,.
                      \een
The statement for the particular solution $q_3(\tau)$ is proved in a similar way.
          \qed

                Denote
                 \be\label{not}
                         \hat{\phi}(\tau) &=& \hat{\phi}_1(\tau) + \hat{\phi}_2(\tau),\nonumber\\[0.15ex]
                         \hat{\varphi}(\tau)  &=& \frac{5+i\,\sqrt{2}}{432}\hat{\varphi}_1(\tau) +
                       \frac{5-i\,\sqrt{2}}{432}\hat{\varphi}_2(\tau),\\[0.2ex]
                       \hat{\psi}(\tau)    & =&
                  \frac{5-i\,\sqrt{2}}{432}\hat{\psi}_1(\tau) + \frac{5+i\,\sqrt{2}}{432}\hat{\psi}_2(\tau)\,.\nonumber
                   \ee

          Now we can build a local formal fundamental matrix solution of the system \eqref{s5} at $\tau=0$.
                     \bth{t1}
                Assume that  $\alpha_3=\frac{1}{8}$. Then the system \eqref{s5} possesses an unique formal  fundamental matrix solution $\hat{\Phi}(\tau)$ at the
              origin in the form
                         $$\,
                      \hat{\Phi}(\tau)=\hat{H}(\tau)\,\tau^{\Lambda}\,\exp\left(\frac{Q}{\tau}\right)\,,
                          \,$$
           where
                      $$\,
                          \Lambda=\diag \left(-\frac{1}{2}, -\frac{1}{2}, -1, -1, -1\right),\quad
                             Q=\diag\left(-2, 2, 0, -4\,i\,\sqrt{2}, 4\,i\,\sqrt{2}\right)\,.
                          \,$$
                  The matrix $\hat{H}(\tau)$ is given by
                  \ben
                      \hat{H}(\tau)=\left(\begin{array}{ccccc}
                               1         & 1    & -\frac{\tau}{16}\,\hat{\phi}(\tau)   & -\frac{i\,\sqrt{2}\,\tau}{6} + \tau\,\hat{\varphi}(\tau)     & \frac{i\,\sqrt{2}\,\tau}{6} + \tau\,\hat{\psi}(\tau)\\
                               -\frac{\tau}{4}        & \frac{\tau}{4}         & h_{23}(\tau)                 & h_{24}(\tau)                 & h_{25}(\tau)\\
                                0        & 0        & 1                  & 1         & 1\\
                                0        & 0        & h_{43}(\tau)                  & h_{44}(\tau)                 & h_{45}(\tau)\\
                                0        & 0        & h_{53}(\tau)                 & h_{54}(\tau)                  & h_{55}(\tau)
                                              \end{array}\right)=
                                      \left(\begin{array}{cc}
                           \hat{H}_1(\tau)               & \hat{H}_{12}(\tau)\\
                                    0                                 & \hat{H}_2(\tau)
                                      \end{array}\right)\,,
                   \een
      where  $\hat{\phi}(\tau), \hat{\varphi}(\tau), \hat{\psi}(\tau)$ are defined by \eqref{not}.

         The entries $h_{i3}(\tau), h_{i4}(\tau), h_{i5}(\tau),\,i=2, 4, 5$   of the matrix $\hat{H}(\tau)$ are defined as follows:

                     \ben
                                 & &
                  h_{23}(\tau)=\frac{\tau^4}{128}\,\hat{\phi}'(\tau)+\frac{\tau^3}{256}\,\hat{\phi}(\tau) + \frac{\tau^2}{8}, \\[0.35ex]
                                & &
                  h_{24}(\tau)= - \frac{\tau^2}{24} + \frac{i\,\sqrt{2}\,\tau^3}{96} - \left(\frac{i\,\sqrt{2}\,\tau^2}{2} + \frac{\tau^3}{16}\right)\,\hat{\varphi}(\tau)
                 - \frac{\tau^4}{8}\,\hat{\varphi}'(\tau),\\[0.35ex]
                              & &
                  h_{25}(\tau)= - \frac{\tau^2}{24} - \frac{i\,\sqrt{2}\,\tau^3}{96} + \left(\frac{i\,\sqrt{2}\,\tau^2}{2} - \frac{\tau^3}{16}\right)\,\hat{\psi}(\tau)
                 - \frac{\tau^4}{8}\,\hat{\psi}'(\tau),\\[0.35ex]
                          & &
                 h_{43}(\tau)=0, \quad h_{44}(\tau)=\frac{i\,\sqrt{2}\,\tau}{4}, \quad h_{45}(\tau)=-\frac{i\,\sqrt{2}\,\tau}{4}, \\[0.35ex]
                            & &
                 h_{53}(\tau)=\frac{\tau^2}{8}, \quad h_{54}(\tau)=-\frac{\tau^2}{8}, \quad h_{55}(\tau)=-\frac{\tau^2}{8}\,.
                  \een

                    \ethe

                     \proof
               When $\alpha_3=\frac{1}{8}$ (resp. $n=0$)
             the system \eqref{s5} admits a local formal fundamental matrix solution $\hat{\Phi}(\tau)$ an the origin in the form
                          $$
                                   \hat{\Phi}(\tau)=\left(\begin{array}{cc}
                                            \hat{\Phi}_1(\tau)               & \hat{\Phi}_{12}(\tau)\\
                                                         0                               & \hat{\Phi}_2(\tau)
                                                              \end{array}\right)\,,
                             $$
                     where $\hat{\Phi}_1(\tau)$ is a local fundamental matrix solution at the origin of the system
                                      \ben
                           q'  &=& -\frac{1}{2 \tau}\,q - \frac{8}{\tau^3} \,p, \\[0.15ex]
          p'  &=& -\frac{1}{2 \tau}\,q +\frac{1 }{2 \tau}\,p.
                               \een
                   In particular,
                     $$\,
                         \hat{\Phi}_1(\tau)=\left(\begin{array}{cc}
                                 \tau^{-1/2}\,e^{-2/\tau}                 & \tau^{-1/2}\,e^{2/\tau}\\[0.15ex]
                             - \frac{\tau^{1/2}\,e^{-2/\tau}}{4}        &   \frac{\tau^{1/2}\,e^{2/\tau}}{4}
                                                       \end{array}\right)\,.
                         \,$$
                    The matrix $\hat{\Phi}_2(\tau)$ is a local fundamental matrix solution at the origin of the system, spanned by the last three equations of the system
                   \eqref{s5}. In particular, taking account of the following choice of $v_1(\tau)=\tau^{-1},\,v_2(\tau)=\tau^{-1}\,e^{-4\,i\,\sqrt{2}/\tau},\,
                               v_3(\tau)= \tau^{-1}\,e^{4\,i\,\sqrt{2}/\tau}$ we find, that
                          $$\,
                               \hat{\Phi}_2(\tau)=\left(\begin{array}{ccc}
                                     v_1(\tau)          & v_2(\tau)              & v_3(\tau)\\
                                      u_1(\tau)         & u_2(\tau)             & u_3(\tau)\\
                                        g_1(\tau)       & g_2(\tau)              & g_3(\tau)
                                                \end{array}\right)=
                          \left(\begin{array}{ccc}
                                    \tau^{-1}                  & \tau^{-1}\,e^{-4\,i\,\sqrt{2}/\tau}                & \tau^{-1}\,e^{4\,i\,\sqrt{2}/\tau}\\[0.15ex]
                                       0                              & \frac{i\,\sqrt{2}}{4}\,e^{-4\,i\,\sqrt{2}/\tau}      & -\frac{i\,\sqrt{2}}{4}\,e^{4\,i\,\sqrt{2}\tau}\\[0.15ex]
                                     \frac{\tau}{8}           & -\frac{\tau}{8}\,e^{-4\,i\,\sqrt{2}/\tau}          & -\frac{\tau}{8}\,e^{4\,i\,\sqrt{2}/\tau}
                                                \end{array}\right)\,.
                                \,$$
                         The matrix $\hat{\Phi}_{12}(\tau)$ has the form
                               \ben
                     \hat{\Phi}_{12}(\tau)=\left(\begin{array}{ccc}
                            q_1(\tau)             & q_2(\tau)       & q_3(\tau)\\[0.15ex]
                             p_1(\tau)            & p_2(\tau)        & p_3(\tau)
                                       \end{array}\right)\,,
                    \een
                       where $q_j(\tau), j=1, 2, 3$ are given by \prref{p1} and $p_j(\tau), j=1, 2, 3$ are computed by
                         $$\,
                         p_j(\tau)=\frac{\tau^3}{8}\left[-\frac{1}{\tau}\,q_j(\tau) - q'_j(\tau) + \frac{1}{\tau}\,v_j(\tau)\right]
                        \,$$
       under the above choice of the solutions $v_j(\tau), j=1, 2, 3$.
               Now the statement is a direct application of the theorem of Hukuhara-Turrittin-Wasov \cite{W}.

           \qed

                  \bpr{p2}
          With respect to the formal fundamental matrix solution $\hat{\Phi}(\tau)$ at the origin, built by \thref{t1} , the formal monodromy $\hat{M}_0$
    and the exponential torus $\T$ of the system \eqref{s5} are given by
                      \ben
      \hat{M}_0=\left(\begin{array}{rrccc}
                   -1      & 0   & 0   & 0    & 0\\
                    0       & -1  & 0   & 0    & 0\\
                   0        & 0   & 1    & 0    & 0\\
                   0        & 0    & 0    & 1    & 0\\
                    0       & 0    & 0    & 0    & 1
                        \end{array}\right),\quad
                 \T=\left\{\left(\begin{array}{ccccc}
                            c_1    & 0   & 0   & 0   & 0\\
                            0        & c_1^{-1}   & 0   & 0   & 0\\
                            0         & 0     & 1    & 0   & 0\\
                            0      & 0     & 0     & c_2   & 0\\
                            0      & 0     & 0     & 0       & c_2^{-1}
                               \end{array}\right),\quad
                        c_1, c_2\in\CC^*\right\}\,.
                         \een
                \epr

         The next lemma provides the 1-sums of the power series $\hat{\phi}_j(\tau), \hat{\varphi}_j(\tau), \hat{\psi}_j(\tau), j=1, 2$ defined by
     \prref{p1}.
                   \ble{sum}
                For any direction $\theta \neq 0$ the function
               $$\,
                 \phi_{1, \theta}(\tau)=\int_0^{+\infty\,e^{i \theta}}
                     \left(1-\frac{\xi}{2}\right)^{-3/2}\,e^{-\frac{\xi}{\tau}}\,d \xi
                \,$$
          defines the 1-sum of the power series $\hat{\phi}_1(\tau)$ in such a direction. Similarly,  for any direction $\theta \neq \pi$ the function
                           $$\,
                         \phi_{2, \theta}(\tau)=
                  \int_0^{+\infty\,e^{i \theta}}
                     \left(1+\frac{\xi}{2}\right)^{-3/2}\,e^{-\frac{\xi}{\tau}}\,d \xi
                    \,$$
                       defines the 1-sum of the power series $\hat{\phi}_2(\tau)$ in such a direction.

                Next, for any  direction $\theta \neq \arg (-4\,i\,\sqrt{2}+2)$ the function
               $$\,
                 \varphi_{1, \theta}(\tau)= \int_0^{+\infty\,e^{i \theta}}
                     \left(1-\frac{\xi}{2-4\,i\,\sqrt{2}}\right)^{-3/2}\,e^{-\frac{\xi}{\tau}}\,d \xi
                 \,$$
           defines the 1-sum of the power series $\hat{\varphi}_1(\tau)$ in such a direction. Similarly,
           for any  direction $\theta \neq \arg (-4\,i\,\sqrt{2}-2)$ the function
                      $$\,
                  \varphi_{2, \theta}(\tau)=
                  \int_0^{+\infty\,e^{i \theta}}
                     \left(1+\frac{\xi}{2+4\,i\,\sqrt{2}}\right)^{-3/2}\,e^{-\frac{\xi}{\tau}}\,d \xi
                    \,$$
                       defines the 1-sum of the power series $\hat{\varphi}_2(\tau)$ in such a direction.

                Finally, for any  direction $\theta \neq \arg (4\,i\,\sqrt{2}+2)$ the function
               $$\,
                 \psi_{1, \theta}(\tau)= \int_0^{+\infty\,e^{i \theta}}
                     \left(1-\frac{\xi}{2+4\,i\,\sqrt{2}}\right)^{-3/2}\,e^{-\frac{\xi}{\tau}}\,d \xi
                \,$$
                defines the 1-sum of the power series $\hat{\psi}_1(\tau)$ in such a direction. Similarly,
                for any  direction $\theta \neq \arg (4\,i\,\sqrt{2}-2)$ the function
                     $$\,
                     \psi_{2, \theta}(\tau)=
                   \int_0^{+\infty\,e^{i \theta}}
                     \left(1+\frac{\xi}{2-4\,i\,\sqrt{2}}\right)^{-3/2}\,e^{-\frac{\xi}{\tau}}\,d \xi
                    \,$$
                       defines the 1-sum of the power series $\hat{\psi}_2(\tau)$ in such a direction.
                  \ele

                  \proof

                             Let $a\in\{2, 2-4\,i\,\sqrt{2}, 2 + 4\,i\,\sqrt{2}\}$.
                   Since $(2 k-1)!! < (2 k)!!=2^k\,k!$ then
                     $$\,
                   \left|\frac{(-1)^{k-1}\,(2 k-1)!!}{2^{k-1}\,a^{k-1}}\right| < \frac{2^k\,k!}{2^{k-1}\,|a|^{k-1}}= \frac{2\,|a|}{|a|^k}\,k!.
                   \,$$
                       Hence the  power series $\hat{\phi}_j(\tau), \hat{\varphi}_j(\tau), \hat{\psi}_j(\tau), j=1, 2$ are Gevrey-1  series with constants $C=4,\,A=1/2$ for the series $\hat{\phi}_j(\tau), j=1, 2$ and
                   $C=12,\, A=1/6$ for the power series $\hat{\varphi}_j(\tau), \hat{\psi}_j(\tau), j=1, 2$. Next, the corresponding formal Borel transforms
                   $\hat{B}_1 \hat{\phi}_j(\xi), j=1, 2$ converge in the open disk $|\xi| < 2$, while $\hat{B}_1 \hat{\varphi}_j(\xi), \hat{B}_1 \hat{\psi}_j (\xi), j=1, 2$
                   converge in the open disk $|\xi| < 6$.  There, we find that
                         \ben
                        \hat{\B}_1 \hat{\phi}_1(\xi)   &=&  \sum_{k=0}^{\infty} \frac{(2 k+1)!!}{2^k}\,\frac{\xi^k}{2^k\,k!}=\left(1-\frac{\xi}{2}\right)^{-3/2}=\phi_1(\xi),\\[0.35ex]
                         \hat{\B}_1 \hat{\phi}_2(\xi)   &=&  \sum_{k=0}^{\infty} (-1)^k\,\frac{(2 k+1)!!}{2^k}\,\frac{\xi^k}{2^k\,k!}=\left(1+\frac{\xi}{2}\right)^{-3/2}=\phi_2(\xi),\\[0.35ex]
                        \hat{\B}_1 \hat{\varphi}_1(\xi)   &=&  \sum_{k=0}^{\infty} \frac{(2 k+1)!!}{2^k}\,\frac{\xi^k}{(2-4 \,i\,\sqrt{2})^k\,k!}=\left(1-\frac{\xi}{2-4\,\i\,\sqrt{2}}\right)^{-3/2}=\varphi_1(\xi),\\[0.35ex]
                         \hat{\B}_1 \hat{\varphi}_2(\xi)   &=&  \sum_{k=0}^{\infty} (-1)^k\,\frac{(2 k+1)!!}{2^k}\,\frac{\xi^k}{(2+4\,i\,\sqrt{2})^k\,k!}=\left(1+\frac{\xi}{2+4\,i\,\sqrt{2}}\right)^{-3/2}=\varphi_2(\xi),\\[0.35ex]
                          \hat{\B}_1 \hat{\psi}_1(\xi)   &=&  \sum_{k=0}^{\infty} \frac{(2 k+1)!!}{2^k}\,\frac{\xi^k}{(2+4\,i\,\sqrt{2})^k\,k!}=\left(1-\frac{\xi}{2+4\,i\,\sqrt{2}}\right)^{-3/2}=\psi_1(\xi),\\[0.35ex]
                         \hat{\B}_1 \hat{\phi}_1(\xi)   &=&  \sum_{k=0}^{\infty} (-1)^k\,\frac{(2 k+1)!!}{2^k}\,\frac{\xi^k}{(2-4\,i\,\sqrt{2})^k\,k!}=\left(1+\frac{\xi}{2-4\,i\,\sqrt{2}}\right)^{-3/2}=\psi_2(\xi)\,.
                          \een
                           Denote $\theta:=\arg (\xi)$. The functions $\phi_j(\xi), \varphi_j(\xi), \psi_j(\xi), j=1, 2$ are continued analytically along any ray $\theta$ from $0$ to $+\infty\,e^{i\,\theta}$, except for
                        $\theta=0, \theta=\pi, \theta=\arg (-4\,i\,\sqrt{2}+2), \theta=\arg (-4\,i\,\sqrt{2}-2), \theta=\arg (4\,i\,\sqrt{2}+2), \theta=\arg (4\,i\,\sqrt{2}-2)$, respectively.

                         Consider the functions
                              \be\label{om}\,
                            \omega_1(\xi)=\left(1 + \frac{\xi}{a}\right)^{-3/2} \quad \textrm{and} \quad
                        \omega_2(\xi)=\left(1 - \frac{\xi}{a}\right)^{-3/2}\,.
                             \ee
                                 Denote $\theta_1:=\arg (a)$. We have that
                                          \ben
                                          |\omega_1(\xi)| \leq \left\{ \begin{array}{ccc}
                                                     1,    & \textrm{if}     & \theta-\theta_1\in \left[-\frac{\pi}{2}, \frac{\pi}{2}\right],\\[0.15ex]
                                                     1/|\sin(\theta-\theta_1)|^{3/2},              &\textrm{if}               & \theta-\theta_1\in \left(\frac{\pi}{2}, \frac{3 \pi}{2}\right)
                                                                                   \end{array}\right.
                                          \een
                                  and
                                          \ben
                                          |\omega_2(\xi)| \leq \left\{ \begin{array}{ccc}
                                                     1,    & \textrm{if}     & \theta-\theta_1\in \left(\frac{\pi}{2}, \frac{3\pi}{2}\right),\\[0.15ex]
                                                     1/|\sin(\theta-\theta_1)|^{3/2},              &\textrm{if}               & \theta-\theta_1\in \left[-\frac{\pi}{2}, \frac{\pi}{2}\right].
                                                          \end{array}\right.
                                          \een
                           Therefore, the Laplace
                   transforms $(\L_1 \omega_1)(\tau)$  and $(\L_1 \omega_2) (\tau)$ are well defined along any ray $\theta \neq \arg (-a)$ and $\theta \neq \arg (a)$, respectively,  from $0$ to $+\infty\,e^{i \theta}$.
              Moreover, the above estimates for $|\omega_j(\xi)|, j=1, 2$
                     ensure that the Laplace transforms
                       \ben
                        \phi_{1, \theta}(\tau)   &=&
                               \int_0^{+\infty\,e^{i \,\theta}}
                               \left(1-\frac{\xi}{2}\right)^{-3/2}\,e^{-\frac{\xi}{\tau}}\,d \xi,\quad
                           \phi_{2, \theta}(\tau)=
                                \int_0^{+\infty\,e^{i \,\theta}}
                               \left(1+\frac{\xi}{2}\right)^{-3/2}\,e^{-\frac{\xi}{\tau}}\,d \xi,\\[0.5ex]
                    \varphi_{1, \theta}(\tau)  &=&
                           \int_0^{+\infty e^{i \theta}}
                      \left(1 - \frac{\xi}{2-4\,i\,\sqrt{2}}\right)^{-3/2}\,e^{-\frac{\xi}{\tau}}\,d \xi,\\[1,5ex]
                         \varphi_{2, \theta}(\tau) &=&
                           \int_0^{+\infty e^{i \theta}}
                      \left(1 + \frac{\xi}{2+4\,i\,\sqrt{2}}\right)^{-3/2}\,e^{-\frac{\xi}{\tau}}\,d \xi,\\[1.5ex]
                             \psi_{1, \theta}(\tau)  &=&
                           \int_0^{+\infty e^{i \theta}}
                      \left(1 - \frac{\xi}{2+4\,i\,\sqrt{2}}\right)^{-3/2}\,e^{-\frac{\xi}{\tau}}\,d \xi,\\[1.5ex]
                                  \psi_{2, \theta}(\tau) &=&
                           \int_0^{+\infty e^{i \theta}}
                      \left(1 + \frac{\xi}{2-4\,i\,\sqrt{2}}\right)^{-3/2}\,e^{-\frac{\xi}{\tau}}\,d \xi
                         \een
                      define  holomorphic functions in the open sector with opening $\pi$
                                  $$\,
                               \mathcal{D}=\left\{ \tau\in\CC^*\,|\,\textrm{Re} \left(\frac{e^{i \,\theta}}{\tau}\right) > 0\right\}
                                    \,$$
            for every ray $\theta \neq \{0, \pi, \arg (-4 i\,\sqrt{2}+2), \arg (-4 \,i\,\sqrt{2} -2), \arg(4\,i\,\sqrt{2}+2), \arg (4\,i\,\sqrt{2}-2)\}$, respectively.

                \qed

                      From \leref{sum} it follows that the sytem \eqref{s5} has 6 singular directions $\theta_1=0, \theta_2=\pi, \theta_3=\arg (2-4 i\,\sqrt{2}),\,\theta_4=\arg (-2-4 i\,\sqrt{2}),
                     \theta_5=\arg (2+4 i\,\sqrt{2}),\,\theta_6=\arg (-2+4 i\,\sqrt{2})$.

                     \bre{r1}
                     Denote by $\tilde{\CC}$  the Riemann surface of the natural logarithm. Let as above
                    $a\in \{2, 2-4\,i\,\sqrt{2}, 2+4\,i\,\sqrt{2}\}$. Denote also by $I=\left(\arg (a), \arg (a) + 2 \pi\right)\subset \RR$ and
                    $J=\left(\arg (-a), \arg(-a) + 2 \pi\right) \subset \RR$ the sets of allowed directions of summation of the power series
                   $\hat{\phi}_1(\tau), \hat{\varphi}_1(\tau), \hat{\psi}_1(\tau)$ and    $\hat{\phi}_2(\tau), \hat{\varphi}_2(\tau), \hat{\psi}_2(\tau)$, respectively.
                  When we move the direction $\theta\in I$ (resp. $\theta\in J$), the holomorphic functions
                 $\phi_{1, \theta}(\tau), \varphi_{1, \theta}(\tau), \psi_{1, \theta}(\tau)$ (resp.     $\phi_{2, \theta}(\tau), \varphi_{2, \theta}(\tau), \psi_{2, \theta}(\tau)$)
                        glue together analytically and define functions $\tilde{\phi}_1(\tau), \tilde{\varphi}_1(\tau), \tilde{\psi}_1(\tau)$
                  (resp.  $\tilde{\phi}_2(\tau), \tilde{\varphi}_2(\tau), \tilde{\psi}_2(\tau)$), which are holomorphic functions on the sector
                             $$\,
                             \tilde{D}_1=\left\{\tau \in\tilde{\CC} \,|\, \arg (a) - \frac{\pi}{2} < \arg (\tau) < \arg(a) + 2 \pi + \frac{\pi}{2}\right\}
                              \,$$
                 (resp.
                             $$\,
                             \tilde{D}_2=\left\{\tau \in\tilde{\CC} \,|\, \arg (-a) - \frac{\pi}{2} < \arg (\tau) < \arg(-a) + 2 \pi + \frac{\pi}{2}\right\})\,.
                              \,$$
                    On these sectors, the functions $\tilde{\phi}_j(\tau), \tilde{\varphi}_j(\tau), \tilde{\psi}_j(\tau)$ are asymptotic to the power series
                 $\hat{\phi}_j(\tau), \hat{\varphi}_j(\tau), \hat{\psi}_j(\tau), j=1, 2$, respectively, in Gevrey 1-sense and define their 1-sums there. The sectors
                 $\tilde{D}_1$ and $\tilde{D}_2$ are the widest sector where 1-Gevrey asymptotic remains valid. The restriction of the function
                 $\tilde{\phi}_j(\tau), \tilde{\varphi}_j(\tau), \tilde{\psi}_j(\tau), j=1, 2$ on $\CC^*$ is a multivalued function. It has only one value
                    $\phi_{j, \theta}(\tau), \varphi_{j, \theta}(\tau), \phi_{j, \theta}(\tau)$ on the sector with opening $\pi$ and bisected by $\theta=\arg(-a)$ for $j=1$
               and $\theta=\arg (a)$ for $j=2$. On the sector with opening $\pi$ and bisected by the corresponding
        singular direction $\theta$   the restriction
                of $\tilde{\phi}_j(\tau), \tilde{\varphi}_j(\tau), \tilde{\psi}_j(\tau)$ has two different values :
                 $\phi^{+}_{j, \theta}(\tau)=\phi_{j, \theta+\epsilon}(\tau)$ and $\phi^{-}_{j, \theta}(\tau)=\phi_{j, \theta+\epsilon}(\tau), j=1, 2$
                      (resp. $\varphi^{+}_{j, \theta}(\tau)=\varphi_{j, \theta+\epsilon}(\tau)$ and $\varphi^{-}_{j, \theta}(\tau)=\varphi_{j, \theta+\epsilon}(\tau)$,
                $\psi^{+}_{j, \theta}(\tau)=\psi_{j, \theta+\epsilon}(\tau)$ and $\psi^{-}_{j, \theta}(\tau)=\phi_{j, \theta+\epsilon}(\tau), j=1, 2$)  for a small number $\epsilon > 0$.
                 \ere

               Thanks to \leref{sum} and \reref{r1} we build an actual fundamental matrix solution at the origin of the system \eqref{s5}, which is associated with the
        formal fundamental matrix solution $\hat{\Phi}(\tau)$ form \thref{t1}. Denote $F(\tau)=\tau^{\Lambda}\,\exp\left(\frac{Q}{\tau}\right)$.
              Denote also, as above,

                 \be\label{not1}
                         \hat{\phi}_{\theta}(\tau) &=& \hat{\phi}_{1, \theta}(\tau) + \hat{\phi}_{2, \theta}(\tau),\nonumber\\[0.15ex]
                         \hat{\varphi}_{\theta}(\tau)  &=& \frac{5+i\,\sqrt{2}}{432}\hat{\varphi}_{1, \theta}(\tau) +
                       \frac{5-i\,\sqrt{2}}{432}\hat{\varphi}_{2, \theta}(\tau),\\[0.2ex]
                       \hat{\psi}_{\theta}(\tau)    & =&
                  \frac{5-i\,\sqrt{2}}{432}\hat{\psi}_{1, \theta}(\tau) + \frac{5+i\,\sqrt{2}}{432}\hat{\psi}_{2, \theta}(\tau)\,.\nonumber
                   \ee

               \bth{act}
                      Assume that $\alpha=\frac{1}{8}$. Then for every nonsingular direction $\theta$ the system \eqref{s5} admits a unique actual fundamental matrix solution
                     $\Phi_{\theta}(\tau)$ at the origin in the form
                             \be\label{afms}
                       \Phi_{\theta}(\tau)=H_{\theta}(\tau)\,F_{\theta}(\tau)\,,
                              \ee
          where $F_{\theta}(\tau)$ is the branch of the matrix $F(\tau)$ for $\theta=\arg (\tau)$. The matrix $\hat{H}_{\theta}(\tau)$  is given by
                       $$\,
                           \hat{H}_{\theta}(\tau)=\left(\begin{array}{cc}
                                H_1(\tau)         & (H_{12}(\tau))_{\theta}\\
                                    0                   & H_2(\tau)
                                      \end{array}\right)\,,
                           \,$$
                   where $H_j(\tau)=\hat{H}_j(\tau), j=1, 2$. The matrix $(H_{12}(\tau))_{\theta}$ is defined as
                      \ben
                       (H_{12}(\tau))_{\theta}=\left(\begin{array}{ccc}
                      -\frac{\tau}{16}\,\phi_{\theta}(\tau)    & -\frac{i\,\sqrt{2}\,\tau}{6} + \tau\,\varphi_{\theta}(\tau)     & \frac{i\,\sqrt{2}\,\tau}{6} + \tau\,\psi_{\tau}(\tau)\\[0.15ex]
                                 (h_{23}(\tau))_{\theta}                 & (h_{24}(\tau))_{\theta}                 & (h_{25}(\tau))_{\theta}
                                          \end{array}\right)\,,
                          \een
         where $\phi_{\theta}(\tau), \varphi_{\theta}(\tau), \psi_{\theta}(\tau)$ are given by \eqref{not1}. The entries $(h_{2j}(\tau))_{\theta}, j, 3, 4, 5$ of the matrix
        $(\hat{H}_{12}(\tau))_{\theta}$ are defined as follows

                     \ben
                                 & &
                  h_{23}(\tau)=\frac{\tau^4}{128}\,\hat{\phi}'_{\theta}(\tau)+\frac{\tau^3}{256}\,\hat{\phi}_{\theta}(\tau) + \frac{\tau^2}{8}, \\[0.35ex]
                                & &
                  h_{24}(\tau)= - \frac{\tau^2}{24} + \frac{i\,\sqrt{2}\,\tau^3}{96} - \left(\frac{i\,\sqrt{2}\,\tau^2}{2} + \frac{\tau^3}{16}\right)\,\hat{\varphi}_{\theta}(\tau)
                 - \frac{\tau^4}{8}\,\hat{\varphi}'_{\theta}(\tau),\\[0.35ex]
                              & &
                  h_{25}(\tau)= - \frac{\tau^2}{24} - \frac{i\,\sqrt{2}\,\tau^3}{96} + \left(\frac{i\,\sqrt{2}\,\tau^2}{2} - \frac{\tau^3}{16}\right)\,\hat{\psi}_{\theta}(\tau)
                 - \frac{\tau^4}{8}\,\hat{\psi}'_{\theta}(\tau)\,.
                    \een

             Near the singular directions $\theta_j, j=1, \ldots, 6$, the system \eqref{s5} admits two different actual fundamental matrix solutions at the origin
                       $$\,
                        \Phi_{\theta_j+\epsilon}(\tau) \quad \textrm{and} \quad \Phi_{\theta_j-\epsilon}(\tau)\,,
                       \,$$
              where $\Phi_{\theta_j \pm \epsilon}(\tau)$ are defined \eqref{afms} for a small number $\epsilon > 0$.
              \ethe

                  In the next theorem we compute explicitly the  Stokes matrices $St_{\theta_j}, j=1, \ldots, 6$.
                  \bth{stokes}
             With respect to the actual fundamental matrix solution at the origin given by \thref{t1}, the system \eqref{s5} has  Stokes matrices $St_0$ and $St_{\pi}$ in the form
                  $$\,
                      St_0=\left(\begin{array}{ccccc}
                       1     & 0   & \mu_1   & 0  & 0\\
                       0     & 1   & 0             & 0  & 0\\
                       0     & 0   & 1             & 0   & 0\\
                       0     & 0   & 0             & 1   & 0\\
                       0     & 0   & 0             & 0   & 1
                                        \end{array}\right),\quad
                      St_{\pi}=\left(\begin{array}{ccccc}
                       1     & 0   & 0   & 0  & 0\\
                       0     & 1   & \mu_2             & 0  & 0\\
                       0     & 0   & 1             & 0   & 0\\
                       0     & 0   & 0             & 1   & 0\\
                       0     & 0   & 0             & 0   & 1
                                        \end{array}\right)\,,
                     \,$$
                       where
                        $$
                                 \mu_1= \frac{i\,2^{3/2}\,\sqrt{\pi}}{4}, \quad    \mu_2= \frac{i\,(-2)^{3/2}\,\sqrt{\pi}}{4}\,.
                          $$
                 Similarly,  with respect to the actual fundamental matrix solution at the origin given by \thref{t1}, the system \eqref{s5} has  Stokes matrices $St_{\theta_3}$ and $St_{\theta_4}$ in the form

                        $$\,
                      St_{\theta_3}=\left(\begin{array}{ccccc}
                       1     & 0   & 0             & \mu_3  & 0\\
                       0     & 1   & 0             & 0  & 0\\
                       0     & 0   & 1             & 0   & 0\\
                       0     & 0   & 0             & 1   & 0\\
                       0     & 0   & 0             & 0   & 1
                                        \end{array}\right),\quad
                      St_{\theta_4}=\left(\begin{array}{ccccc}
                       1     & 0   & 0   & 0  & 0\\
                       0     & 1   & 0             & \mu_4  & 0\\
                       0     & 0   & 1             & 0   & 0\\
                       0     & 0   & 0             & 1   & 0\\
                       0     & 0   & 0             & 0   & 1
                                        \end{array}\right)\,,
                     \,$$
                       where
                        $$
                                 \mu_3= \frac{(2 -5\,i\,\sqrt{2})\,(1-2\,i\,\sqrt{2})^{3/2}\,\sqrt{\pi}}{54}, \quad     \mu_4= -\frac{(2 +5\,i\,\sqrt{2})\,(-1-2\,i\,\sqrt{2})^{3/2}\,\sqrt{\pi}}{54}\,.
                          $$

                 Similarly,  with respect to the actual fundamental matrix solution at the origin given by \thref{t1}, the system \eqref{s5} has  Stokes matrices $St_{\theta_5}$ and $St_{\theta_6}$ in the form

                        $$\,
                      St_{\theta_5}=\left(\begin{array}{ccccc}
                       1     & 0   & 0             & 0  & \mu_5\\
                       0     & 1   & 0             & 0  & 0\\
                       0     & 0   & 1             & 0   & 0\\
                       0     & 0   & 0             & 1   & 0\\
                       0     & 0   & 0             & 0   & 1
                                        \end{array}\right),\quad
                      St_{\theta_6}=\left(\begin{array}{ccccc}
                       1     & 0   & 0   & 0  & 0\\
                       0     & 1   & 0             & 0  & \mu_6\\
                       0     & 0   & 1             & 0   & 0\\
                       0     & 0   & 0             & 1   & 0\\
                       0     & 0   & 0             & 0   & 1
                                        \end{array}\right)\,,
                     \,$$
                       where
                        $$
                                 \mu_5= -\frac{(2 +5\,i\,\sqrt{2})\,(1+2\,i\,\sqrt{2})^{3/2}\,\sqrt{\pi}}{54}, \quad     \mu_6= \frac{(2 -5\,i\,\sqrt{2})\,(-1+2\,i\,\sqrt{2})^{3/2}\,\sqrt{\pi}}{54}\,.
                          $$
            \ethe

         \proof

          Let just above   $a\in\{2, 2-4\,i\,\sqrt{2}, 2 + 4\,i\,\sqrt{2}\}$.
      Consider the Laplace  transforms
       $$\,
              \L_1(\omega_1, \theta, a)=\int_0^{+\infty\,e^{i \theta}}
                           \left(1+\frac{\xi}{a}\right)^{-3/2}\,e^{-\frac{\xi}{\tau}}\,d \xi
    \,$$
and
              $$\,
             \L_1(\omega_2, \theta, a)=\int_0^{+\infty\,e^{i \theta}}
                           \left(1-\frac{\xi}{a}\right)^{-3/2}\,e^{-\frac{\xi}{\tau}}\,d \xi
         \,$$
   for $\theta\neq \arg(-a)$ and $\theta \neq \arg(a)$, respectively, of the functions $\omega_1(\xi)$ and $\omega_2(\xi)$ from \eqref{om}.

           Let $\epsilon > 0$ be a small number. Let $\theta-\epsilon$ and $\theta+\epsilon$ be two nonsingular neighboring directions of the singular direction $\theta$.
             Comparing $\L_1(\omega_2, \theta+\epsilon, a)$ and $\L_1(\omega_2, \theta-\epsilon, a)$, we find that
                   $$\,
               \L_1(\omega_2, \theta-\epsilon, a)-\L_1(\omega_2, \theta+\epsilon, a)=\int_{\gamma}
                \left(1-\frac{\xi}{a}\right)^{-3/2}\,e^{-\frac{\xi}{\tau}}\,d \xi\,,
                  \,$$
           where $\gamma=(\theta-\epsilon) - (\theta+\epsilon)$. Without changing the integral, we can deform the path $\gamma$ into a Hankel-type contour $\gamma_1$,
                      going along $\theta=\arg (a)$, starting from $+\infty\,e^{i\,\arg(a)}$, encircling $a$ in the positive sense and returning to $+\infty\,e^{i\,\arg(a)}$. The change
      $1-\frac{\xi}{a}=-\frac{\eta}{a}$ takes the contour $\gamma_1$ into a Hankel-type contour $\gamma_2$ along $\theta=\arg(a)$ from $+\infty\,e^{i \,\arg(a)}$, encircling
 $0$ in the positive sense and returning to $+\infty\,e^{i\,\arg (a)}$. Then
                   $$\,
               \L_1(\omega_2, \theta-\epsilon, a)-\L_1(\omega_2, \theta+\epsilon, a)=e^{-a/\tau}\,\int_{\gamma_2}
                \left(-\frac{\eta}{a}\right)^{-3/2}\,e^{-\frac{\eta}{\tau}}\,d \eta\,.
                  \,$$
     Next, the transformation $\eta/\tau=-\beta$ takes the contour $\gamma_2$ into a Hankel-type contour $\gamma_3$, which winds around $\RR^{-}$, starting from $-\infty$,
 encircling $0$ in the positive sense and returning to $-\infty$. Then
                                   $$\,
                \L_1(\omega_2, \theta-\epsilon, a)-\L_1(\omega_2, \theta+\epsilon, a)=-a^{3/2}\,\tau^{-1/2}\,e^{-a/\tau}\,\int_{\gamma_3}
                \beta^{-3/2}\,e^{\beta}\,d \beta\,.
                  \,$$
          The last contour integral is the Hankel's representation of the reciprocal Gamma function $1/\Gamma(m),\,m\neq 0, -1, -2, \ldots$
                   $$\,
                      \frac{1}{\Gamma(m)}=\frac{1}{2 \,\pi\,i}\,\int_{\gamma_3} e^{\beta}\,\beta^m\,d \beta, \qquad |\arg (\beta)| \leq \pi.
                     \,$$
       As a result, we find that
                                   \be\label{om1}
                \L_1(\omega_2, \theta-\epsilon, a)-\L_1(\omega_2, \theta+\epsilon, a)=-4\,i\,\sqrt{\pi}\,a^{3/2}\,\tau^{-1/2}\,e^{-a/\tau}\,.
                  \ee

                  In the same manner, one can compute that
                                                       \be\label{om2}
                \L_1(\omega_1, \theta-\epsilon, a)-\L_1(\omega_1, \theta+\epsilon, a)=-4\,i\,\sqrt{\pi}\,(-a)^{3/2}\,\tau^{-1/2}\,e^{a/\tau}\,.
                  \ee

       In order to compute the multiplier $\mu_1$ we have to compare the solutions $[q_1(\tau)]^{-}_0$ and $[q_1(\tau)]^{+}_0$, corresponding to the directions
              $0-\epsilon$ and $0+\epsilon$, respectively. Then for $\textrm{Re}(\tau) > 0$ applying the formula \eqref{om1}, we find  that
                \ben
            [q_1(\tau)]^{-}_0 - [q_1(\tau)]^{+}_0   &=&
                            -\frac{1}{16} \left[\L_1(\omega_2, 0-\epsilon, 2) - \L_1(\omega_2, 0+\epsilon, 2)\right]
                  =\frac{i\,2^{3/2}\,\sqrt{\pi}}{4}\,\tau^{-1/2}\,\exp\left(-\frac{2}{\tau}\right)\\[0.15ex]
                                                       &=&
                     \mu_1\,\tau^{-1/2}\,\exp\left(-\frac{2}{\tau}\right)\,.
                  \een
                 In the same manner, for $\textrm{Re}(\tau) < 0$ applying the formula \eqref{om2}, we find that
                \ben
            [q_1(\tau)]^{-}_{\pi} - [q_1(\tau)]^{+}_{\pi}   &=&
                     -\frac{1}{16} \left[\L_1(\omega_1, \pi-\epsilon, 2) - \L_1(\omega_1, \pi+\epsilon, 2)\right]
                  =\frac{i\,(-2)^{3/2}\,\sqrt{\pi}}{4}\,\tau^{-1/2}\,\exp\left(\frac{2}{\tau}\right)\\[0.15ex]
                               &=&
                     \mu_2\,\tau^{-1/2}\,\exp\left(\frac{2}{\tau}\right)\,.
                  \een

                  In the same way, applying the formulas \eqref{om1} and \eqref{om2} one can compute the Stokes multipliers $\mu_j, j=3, 4, 5, 6$.
             \qed

              Now we are in a position to describe the differential Galois group of the system \eqref{s5}.

               \bth{gs5}
                  Assume that $\alpha_3=\frac{1}{8}$. Then the connected component of the unit element of the differential Galois group of the system \eqref{s5} is not Abelian.
                  \ethe

                   \proof
            The local differential Galois group at $\tau=\infty$ of the system \eqref{s5} can be interpreted as a subgroup of the local differential Galois group
           at the origin. Hence if  we  prove that the connected component of the local differential Galois group at the origin is not Abelian and so be the
             connected component of the whole differential Galois group.

             From  \prref it follows that the group generated by the formal monodromy and the exponential torus is not a connected group.
              However, the connected component of the unit element of this group coincides with the exponential torus $\T$. Then from the theorem of Ramis it
                follows that the local differential Galois group at the origin of the system \eqref{s5} is generated topologically by the exponential torus $\T$
                 and the Stokes matrices $St_{\theta_j}, j=1, \ldots,  6$. Denote by $S$ the Zariski closure of the subgroup generated by the Stokes matrices. Then
               the element $S_{\nu}$ of $S$ has the form
                    $$\,
                S_{\nu}=\left(\begin{array}{ccccc}
                                1   & 0    & \nu_1  & \nu_3  &\nu_5\\
                                0   & 1    & \nu_2  & \nu_4   & \nu_6\\
                                0   & 0    & 1          & 0           & 0\\
                                0   & 0    & 0          & 1           & 0\\
                                0   & 0    & 0          & 0            & 1
                                    \end{array}\right)\,,
                  \,$$
               where $\nu_j\in\CC, j=1, \ldots, 6$.  Denote by $T$ the Zariski closure of the subgroup generated by the exponential torus $\T$. Then the element $T_{c_1, c_2}$
                 of $T$ has the form
                          form
                    $$\,
                T_{c_1, c_2}=\left(\begin{array}{ccccc}
                                c_1   & 0    & 0  & 0  &0\\
                                0   & c_1^{-1}    & 0  & 0   & 0\\
                                0   & 0    & 1          & 0           & 0\\
                                0   & 0    & 0          & c_2           & 0\\
                                0   & 0    & 0          & 0            & c_2^{-1}
                                    \end{array}\right)\,,
                  \,$$
                 where $c_1, c_2\in\CC^*$.

            When $\nu_j\neq 0, j=1, \ldots, 6$ and $c_1\neq 1$, the commutator between $S_{\nu}$ and $T_{c_1, c_2}$
                        $$\,
                         S_{\nu}\,T_{c_1, c_2}\,S^{-1}_{\nu}\,T^{-1}_{c_1, c_2}=\left(\begin{array}{ccccc}
                               1       & 0   & \nu_1 (1-c_1)    & \nu_3 (1-c_1\,c^{-1}_2)     & \nu_5 (1- c_1\,c_2)\\
                               0       & 1   & \nu_2 (1-c^{-1}_1)   & \nu_4 (1-c^{-1}_2 c^{-1}_2)   & \nu_6 (1-c^{-1}_1\,c_2)\\
                               0       & 0   & 1     & 0    & 0\\
                               0        & 0   & 0    & 1    & 0\\
                               0      &0     &0      & 0    & 1
                                        \end{array}\right)
                       \,$$
                   is not identically equal to the identity matrix.

                  The condition $c_1=1$ implies that for every element $\sigma$ of the differential Galois group of the system \eqref{s5} we will have
                  $\sigma (e^{-2/\tau})=e^{-2/\tau}$, which is a contradiction since $e^{-2/\tau}\notin \CC(\tau)$. Thus the connected component of the unit element
                   of the differential Galois group of the system \eqref{s5} is not Abelian.
           \qed

                 \vspace{2ex}

               {\it Proof of \thref{key2}.}\,
              The transformations \eqref{ch} change the differential Galois group of the $(\textrm{LNVE})_2$. The differential Galois group of the reduced system
                \eqref{s5} is isomorphic to the differential Galois group of the $(\textrm{LNVE})_2$. Then combining \thref{gs5} with the Morales-Ramis-Sim\'o theory,
            we obtain that when $(\alpha_0, \alpha_1, \alpha_2, \alpha_3)=\left(\frac{1}{16}, \frac{1}{16}, \frac{1}{8}, \frac{1}{8}\right)$ the Sasano system of type
              $A^{(2)}_5$ is not integrable by rational first integrals.
             \qed

%%%%%%%%%%%%%%%%%%%%%%%%%%%%%%%%%%%%%%%%%%%%%%%%%%%%%%%%%%%%%%%%%%%%
%Backlund
%%%%%%%%%%%%%%%%%%%%%%%%%%%%%%%%%%%%%%%%%%%%%%%%%%%%%%%%%%%%%%%%%%%%%%

             \section{Proof of the \thref{main}}

  In this section, with the aid of the  B\"acklund transformations \eqref{sym}, we will extend the results of the previous two sections and we will establish the main result of this paper.

  Let us first note that if $y\equiv 0$ then $\alpha_0=0$, if $w \equiv 0$ then $\alpha_1=0$, if $y+w-1=0$ then $\alpha_3=0$, and if $x\,z+t=0$ then $\alpha_2=0$.
  This observation allows us to consider the transformations $s_0, s_1, s_2, s_3$ as identically transformations when $y \equiv 0, w \equiv 0, x\,z+t \equiv 0, y+w -1 \equiv 0$,
  respectively.
           \vspace{2ex}

   {\it Proof of \thref{main}}.\,
   Proof follows from \leref{orbit}, \thref{M} and the fact that the B\"acklund transformations \ref{sym} are canonical transformations, which are rational ones in all
  canonical variables.\qed

%%%%%%%%%%%%%%%%%%%%%%%%%%%%%%%%%%%%%%%%%%%%%%%%%%%%%%%%%%%%%%%%%%%%%%%%%%

   \vspace{1cm}

   {\bf Acknowledgments.}\,
   The author was partially supported by Grant 80-10-30 / 21.05.2025  of the Sofia University "St. Kliment Ohridski" Science Foundation.

%%%%%%%%%%%%%%%%%%%%%%%%%%%%%%%%%%%%%%%%%%%%%%%%%%%

 \vspace{1cm}
 {\bf Data availability statement.}\
   No new data were created or analysed during the current study.

%%%%%%%%%%%%%%%%%%%%%%%%%%%%%%%%%%%%%%%%%%%%%%%%%%%%%%%%%%%%%%%%%%
% theory
%%%%%%%%%%%%%%%%%%%%%%%%%%%%%%%%%%%%%%%%%%%%%

	%%%%%%%%%%%%%%%%% References %%%%%%%%%%%%%%%%%%%%%%%%%%%%%%%%%%%%%%%%%%%%%%%
\begin{small}
    
\end{small}
%%%%%%%%%%%%%%%%%%%%%%%%%%%%%%%%%%%%%%%%%%%%%%%%%%%%%%%%%%%%%%%

\end{document}